\documentclass[aps,pre,longbibliography,notitlepage,unsortedaddress,floatfix,nofootinbib]{revtex4-1}

\usepackage[format=plain,labelfont={bf,small},textfont=small,justification=raggedright,singlelinecheck=false]{caption}
\usepackage{bm,amsmath,empheq,amssymb,graphicx,stmaryrd,float,subcaption,upgreek,sidecap,dsfont,mathtools,gensymb,multirow,array}
\usepackage[multiple]{footmisc}
\usepackage[abs]{overpic}
\usepackage{hyperref}
\usepackage{comment,footnote,mathrsfs,subcaption,xcolor}

\newcommand{\jump}[1]{{\left\llbracket #1 \right\rrbracket}} 
\newcommand{\avg}[1]{{\langle #1 \rangle}}

\newcommand{\bL}{{\bf L}}

\newcommand{\bl}{{\bm l}}
\newcommand{\bmo}{{\bm m}}

\newcommand{\bn}{{\bm n}}
\newcommand{\bP}{{\bf P}}

\newcommand{\bE}{{\bf E}}

\newcommand{\bzero}{{\bf 0}}

\newcommand{\bp}{{\bm p}}
\newcommand{\br}{{\bm r}}
\newcommand{\bd}{{\bm d}}

\usepackage[utf8]{inputenc}
\usepackage[T1]{fontenc}
\usepackage{nomencl}

\cleardoublepage           
\renewcommand{\nomname}{} %% may be this is good.
\setcitestyle{numbers,square}
\begin{document}
\title{Self-contact in a buckled elastica}
\author{Krishnan Suryanarayanan}
\email{krishnan@manit.ac.in}
\affiliation{Maulana Azad National Institute of Technology Bhopal, Bhopal, Madhya Pradesh-462003}

\author{Parth Patel}
%\email{parthkumar.patel@iitgn.ac.in}

\author{Anup Kumar Pathak}
%\email{anup.pathak@iitgn.ac.in}

\author{Harmeet Singh}
\thanks{Corresponding author}
\email{harmeet.singh@iitgn.ac.in}
\affiliation{Indian Institute of Technology Gandhinagar, Palaj, Gujarat-382055}
\date{\today}

\begin{abstract}
We explore the mechanics of a terminally loaded buckled elastica under frictionless self-contact.
With the aid of two integrals associated with the elastica, we propose a scale-invariant condition necessary for the onset of contact.
The condition is independent of the boundary conditions, does not involve the position vectors of the material points, and delivers the value of the compressive load at which self-contact initiates.
Furthermore, we show that one of the two integrals, namely the \emph{Hamiltonian}, persists after contact.
We compute post-contact configurations of modes three through ten for a pinned-pinned buckled elastica.
At a given value of the compressive load, we report multiple post-contact configurations for modes eight and nine.
Finally, we show that an infinite force is required to transition from a point contact to a line contact in symmetric post-contact configurations of odd modes.
\end{abstract}
\maketitle

\section{Introduction}\label{sec:introduction}
Consider the configuration of an Euler elastica buckled under compressive loads.
As the loads are increased and the deflections become large, different segments of the elastica may come into contact with one another.
The theory of the Euler elastica enables the computation of the critical load at which the rod buckles, as well as its post-buckling shapes. 
It does not, however, account for the impenetrability of the elastica, and therefore permits different segments to interpenetrate.

In 1973, Flaherty and Keller \cite{flahertykeller1973} considered the problem of a buckled elastica while accounting for self-contact.
They analyzed the first buckling mode of an elastica with fixed-fixed boundary conditions, and the third buckling mode with pinned-pinned boundary conditions. 
For both cases, they identified the critical load at which contact occurs, and derived asymptotic formulas for the location of the contact point post-contact as a function of the compressive load.
They showed that, in both cases, the section of the rod between the contact points is described by a similarity solution.
As the compressive force increases, the points of contact recede toward the respective nearer ends.

Motivated by \cite{flahertykeller1973}, we present a formalism to analyze self-contacting solutions of a planar elastica.
Our approach differs from theirs in two key aspects.
The first is the detection of the critical contact load. 
While Flaherty and Keller~\cite{flahertykeller1973} state the compressive loads (up to five significant digits) at which contact occurs, for both pinned-pinned and fixed-fixed boundary conditions, they do not provide a method, or criterion, to detect the onset of contact.
We can only assume that they employed a trial-and-error scheme to detect contact by comparing the distances between material points in ambient space.
In this work, we present a scale-invariant condition that is necessary (but not sufficient) for the occurrence of self-contact in inflectional solutions of the planar elastica. 
The condition states that the ratio of two integrals associated with the elastica \cite{singh2019}, namely the \emph{Hamiltonian} \cite{steigmannfaulkner1993,oreilly07,singh2022,neukirch2025}, and the magnitude of the internal force, must necessarily reach a critical value for contact to occur.  
This condition does not involve comparing distances between material points in ambient space; furthermore, it does not depend on the boundary conditions, and can be employed to detect the onset of contact in any planar inflectional configuration.
We also show that unlike the conservation of the internal force, the conservation of the Hamiltonian along the length of the rod persists post-contact.

The second aspect in which our work differs from that of \cite{flahertykeller1973} is the numerical technique used to compute post-contact configurations.
Flaherty and Keller~\cite{flahertykeller1973} employ the shooting method to solve the nonlinear boundary value problem governing the post-contact equilibrium of a buckled elastica. 
Shooting methods typically involve guessing and iterating the boundary conditions at one end of the rod so that the solution satisfies the boundary conditions at the other \cite{ascherpetzold1998}. 
Such methods tend to be highly sensitive to initial guesses and perform poorly when applied to stiff boundary value problems. We perform our computations by discretising the boundary value problem using the method of orthogonal collocation, and solve the resulting algebraic equations via pseudo-arclength continuation~\cite{doedeletal1991}, as implemented in \texttt{AUTO-07p}~\cite{doedeldldeman2007AUTO}. 
This approach removes the necessity of providing manual initial guesses; instead, it constructs its own estimates that are guaranteed to converge at each step of the continuation process~\cite{doedeletal1991}, provided the step size is sufficiently small. It typically requires a known base solution, subject to the conditions imposed by the implicit function theorem, for an initial value of the continuation parameter. 
In our case, the problem is formulated with the magnitude of the compressive load as the continuation parameter, and the undeformed straight configuration of the elastica at zero load is used to initiate the continuation process.

% Continuation methods typically require a known solution at some initial value of the continuation parameter.
% We take this solution to be the straight configuration at zero load and, by way of bifurcation analysis, jump onto a buckled mode.
% We then detect the onset of contact by using the scale-invariant condition and use that result as a starting solution to compute post-contact configurations.

We study modes three through ten of the pinned-pinned elastica. 
First, we compute the locations of the material points on the rod that come together at the onset of contact.
We show that the post-contact solutions for odd modes, in general, comprise only two shapes corresponding to the classical Euler elastica.
Modes five, seven, and nine appear to be a modular repetition of mode three.
Even modes, on the other hand, comprise three shapes in general, with mode six being an exception consisting of four shapes.
Mode eight and ten appear to be modular repetitions of mode four.
We also compute multiple post-contact configurations for modes eight and mode nine at a given value of the compressive load.

Our analysis reveals a curious fact that no matter how high the compressive load becomes, the points of contacts never transition into line contact.
Such transitions are fairly common when slender rods are in contact with self~\cite{flaherty1972,coleman2000,chamekh2009}, and with external objects \cite{domokos1997,plaut1999,pocheau2004,chenli2007,huynen2016,grandgeorge2021,singh2022}.
We present a limited theoretical argument showing that point contacts in symmetric configurations of odd modes can never transition into line contacts. 
For even modes, we present numerical evidence supporting the same conclusion.

The particular problem and the associated formalism may be applicable to several scenarios that entail the self-contact of elasticae.
Self-contact appears in problems ranging from the self-encapsulation of elastic rods \cite{bosi2015} to the snapping of beams \cite{chenlee2021,plaut2008}.
Snapping of a fixed-fixed beam, when the relative angle between the two ends are continuously varied, may also result in self-contact for certain pairs of end angles and distances between the two ends \cite{chenlee2021-2}. 
These phenomena may be of relevance to MEMS devices \cite{zhang2015,maurini2007,krylov2008}.
Other areas in which self-contact is relevant include the supercoiling of DNA \cite{coleman2000,shihearst1994}, and the buckling of pressurized tubes aimed at understanding the mechanics of arterial collapse \cite{flaherty1972,han2012}.

The current study could also be of relevance to the optimal packing of deployable antennas for space applications, where estimations of the stored strain energy, and safe ways to deploy them in space without violent expulsion, are of crucial importance \cite{arya2015}.
Following are some relevant articles on self-avoiding elasticae under confinement. 
The mechanics of packing slender bodies in two-dimensional, rigid confined cavities, where self-contact inevitably arises, has been studied experimentally in \cite{Donato2003Scaling}.
An experimental and numerical study regarding a similar problem on the packing of flexible structures was presented in \cite{boue2006spiral}, wherein the authors utilized an elastica model to solve the underlying nonlinear boundary value problems via the shooting method.
Napoli and Goriely \cite{napoli2017Tale} provided analytical solutions governing the symmetric equilibria of two planar nested elastic rings, where the inner ring is longer than the outer ring, by means of a variational formulation.
Asymmetric equilibria of the same problem have been studied numerically in \cite{lombardoetal2018} using a planar elastica model. 
Furthermore, Cutolo et al. \cite{Cutolo2023Growth} have provided analytical solutions for both symmetric and asymmetric equilibria of an intrinsically straight elastic rod growing inside a flexible ring.

The manuscript is organized as follows. In Sec.~\ref{sec:essential_theory}, we lay out the necessary theory of planar elastic rods. Sec.~\ref{sec:euler_elastica} recapitulates the classification of the classical Euler elastica, which forms the basis of the scale-invariant contact condition. In Sec.~\ref{sec:scale_invariant_contact_condition}, we present the aforementioned contact condition. The pinned–pinned buckled elastica is considered in Sec.~\ref{sec:pinned_pinned_buckled_elastica}, with the pre-contact, onset of contact, and post-contact regimes discussed in the subsequent subsections. Post-contact solutions for the even and odd modes of the pinned–pinned buckled elastica are presented in Sec.~\ref{sec:self_contacting_solutions}. We then comment on the impossibility of point contacts transitioning into line contact in Sec.~\ref{sec:line_contact}. Finally, we conclude in Sec.~\ref{sec:conclusion}.

\section{Planar elastic rods}\label{sec:essential_theory}
We represent any configuration of a planar elastic rod as a planar curve $\br\equiv\br(s)$, parametrized by its arc-length coordinate $s$, and two orthonormal directors $\{\bd_1(s),\bd_3(s)\}$ that span the plane of deformation \cite{antman:nonlinear,oreillybook}.
The deformation of the rod is constrained by the following kinematic restrictions,
\begin{align}\label{eq:inextensibility_unshearability}
    \br'=\bd_3\, ,\qquad \bd_1' = -\kappa \bd_3\, ,\qquad \bd_3' =  \kappa \bd_1\, ,
\end{align}
where $\kappa\equiv\kappa(s)$ is the signed Frenet curvature of $\br$.
The first relation enforces the inextensiblity and unshearability of the elastic rod, whereas the other two ensure the orthonormality of the directors as they vary along $s$.

The force and moment balance of a finite segment $[s_1,s_2]$ of a planar elastica can be written as
\begin{subequations}\label{eq:force_moment_balance_integral}
\begin{align}
    [\bn]_{s_1}^{s_2} + \int_{s_1}^{s_2}\!\!\bp~ds &= \bzero\, ,\label{eq:force_balance_integral}\\
    [\bmo + \br\times\bn]_{s_1}^{s_2} + \int_{s_1}^{s_2}\!\!\br\times\bp~ds + \int_{s_1}^{s_2}\!\!\bl~ds &= \bzero\, .\label{eq:moment_balance_integral}
\end{align}
\end{subequations}
Here, $\bn\equiv\bn(s)$ and $\bmo\equiv\bmo(s)$ denote the internal force and moment respectively.
They are defined as the net force and moment exerted by the material at $s^+$ on the material at $s^-$, where $s^\pm = \lim_{\epsilon\rightarrow 0} (s\pm \epsilon)$, with $\epsilon>0$.
The vectors $\bp\equiv\bp(s)$ and $\bl\equiv\bl(s)$ denote, respectively, the external force and moment densities (per unit arc-length) experienced by $\br$.
Assuming sufficient regularity of the fields appearing in \eqref{eq:force_moment_balance_integral}, the two integral balances can be localized to obtain the local equilibrium equations, 
\begin{subequations}\label{eq:force_moment_balance_local}
\begin{align}
    \bn' + \bp &= \bzero\, ,\label{eq:force_balance_local}\\
    \bmo' + \br'\times\bn + \bl &= \bzero\, .\label{eq:moment_balance_local}
\end{align}
\end{subequations}
\begin{figure}[t]
    \centering
    \includegraphics[width=0.5\linewidth]{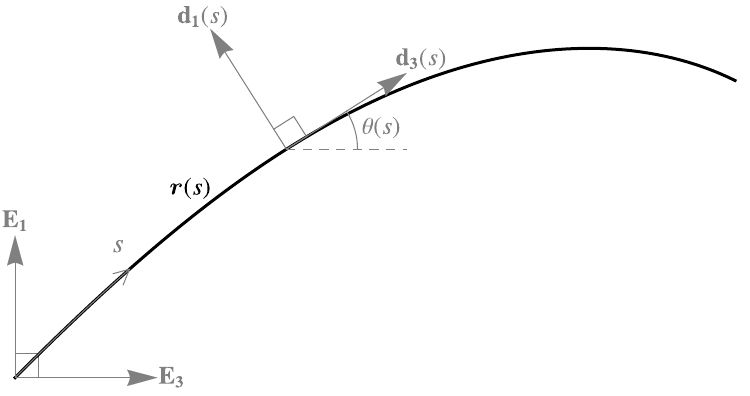}
    \caption{Schematic of a planar elastica. The position vector of a material point, identified by its arc-length coordinate $s$, is denoted by $\br(s)$. The director frame $\{\bd_1(s),\bd_3(s)\}$ differs from the Cartesian frame $\{\bE_1,\bE_3\}$ by an anticlockwise rotation of $\theta\equiv\theta(s)$.}
    \label{fig:schematic}
\end{figure}
Contact problems in inextensible and unshearable elastic rods often admit concentrated forces and moments \cite{burridgekeller1978,oreillybook,hannasingh2018,singh2022}, which can be approximated by the following representation of the external force and moment densities \cite{maddockskeller1987},
\begin{align}\label{eq:point_force_moment}
    \bp = \bP\delta(s-s_0)\, ,\qquad \bl = \bL\delta(s-s_0)\, ,
\end{align}
where $s_0\in[s_1,s_2]$ is the arc-length coordinate of a point of contact, and $\delta(s)$ is the Dirac delta function. 
Assuming the continuity of $\br$ and $\br'$, \eqref{eq:force_moment_balance_integral} can be localized around $s_0$ to obtain the following jump conditions,
\begin{subequations}\label{eq:force_moment_jump_conditions}
    \begin{align}
        \jump{\bn} + \bP & = \bzero\, ,\label{eq:force_jump}\\
        \jump{\bmo} + \bL &= \bzero\, ,\label{eq:moment_jump}
    \end{align}
\end{subequations}
where $\jump{\mathcal{A}}=\mathcal{A}^+ - \mathcal{A^-}$ denotes the jump in a field quantity across $s_0$.

The internal moment and the curvature of a planar elastica are linearly related by the following constitutive relation,
\begin{align}
    \bmo = K\kappa \bd_2\qquad \text{where} \qquad \bd_2:=\bd_3\times\bd_1\, .\label{eq:constitutive_relation}
\end{align}
Here $K$ is the bending modulus given by $K=EI$, with $E$ being Young's modulus and $I$ the second moment of area of the elastica cross-section. 
Also, we note that $\bd_2' = \bzero$ since $\{\bd_1,\bd_3\}$ are confined to the plane of deformation.

A planar elastica loaded only at its two ends, i.e. with $\bp=\bzero$ and $\bl=\bzero$, admits the following two integrals,
\begin{subequations}\label{eq:conservation_laws}
\begin{align}
    \bn(s) &= \bn(0)\, ,\label{eq:conservation_of_force}\\
    H(s) &= H(0)\, ,\label{eq:conservation_of_Hamiltonian}
\end{align}
\end{subequations}
in regions away from points of singularity, i.e. $\forall s\in[s_1,s_2]\setminus\{s_0\}$. 
Here $H\equiv H(s)$ is the \emph{Hamiltonian} \cite{steigmannfaulkner1993,maddocksdichman1994,singh2022,neukirch2025} function defined as
\begin{align}\label{eq:Hamiltonian}
    H = \bn\cdot\bd_3 + \frac{\bmo\cdot\bmo}{2K}\, .
\end{align}

Finally, we introduce a fixed Cartesian frame $\{\bE_1,\bE_3\}$ from which the director frame $\{\bd_1,\bd_3\}$ differs by a counterclockwise rotation of angle $\theta\equiv\theta(s)$ (see Fig~\ref{fig:schematic}).
The position vector and the directors are represented in the Cartesian basis as follows,
\begin{subequations}\label{eq:planar_representations}
\begin{alignat}{3}
    \br     &= r_1\bE_1 + r_3\bE_3,    &\quad \bd_1 &= \cos\theta\bE_1 - \sin\theta\bE_3, &\quad \bd_3 &= \sin\theta\bE_1 + \cos\theta\bE_3, \label{eq:r_d1_d3} \\
    \kappa  &= \theta',                &\quad \bn   &= n_1\bd_1 + n_3\bd_3,               &\quad H     &= n_3 + \frac{1}{2}K\kappa^2. \label{eq:kappa_n_H}
\end{alignat}
\end{subequations}
The first relation in \eqref{eq:kappa_n_H} is obtained by employing the planar representation of the directors above in \eqref{eq:inextensibility_unshearability}$_{2,3}$, while \eqref{eq:kappa_n_H}$_3$ is obtained by using the constitutive relation \eqref{eq:constitutive_relation}, and \eqref{eq:kappa_n_H}$_2$, in \eqref{eq:Hamiltonian}.
\subsection{The nature of contact}\label{sec:nature_of_contact}
The nature of self-contact in the elastica is assumed to be frictionless and non-adhesive, implying that at all contact points the point forces and moments must satisfy,
\begin{align}\label{eq:frictionless_non_adhesive}
    \bP\cdot\bd_3 = \bzero\, ,\qquad \bL=\bzero\, .
\end{align}
Under these conditions, it can be shown that the jump in the Hamiltonian \eqref{eq:Hamiltonian} at all contact points vanishes:
\begin{align}\label{eq:jump_Hamiltonian}
    \jump{H} = -\bP\cdot\bd_3 - \frac{\bL\cdot\avg{\bmo}}{K} = 0\, .
\end{align}
Here we have used the continuity of the tangent across the contact points, along with \eqref{eq:force_jump} and \eqref{eq:moment_jump} to arrive at the first equality.

Equation \eqref{eq:jump_Hamiltonian} along with the conservation law \eqref{eq:conservation_of_Hamiltonian} implies that \emph{the} Hamiltonian for self-contacting configuration is conserved throughout the length of the rod. 
The same is not true of the conservation of the internal force \eqref{eq:conservation_of_force}, which will suffer a jump $-\bP\cdot\bd_1$ across points of contact.

\section{Classical Euler elastica}\label{sec:euler_elastica}
Central to our discussion is a classification of the classical Euler elastica presented by Singh and Hanna in \cite{singh2019}, based on which we propose a scale-invariant condition necessary for the onset of contact.
We briefly review the relevant material from \cite{singh2019} for the reader's convenience.

The classical Euler elastica can be modeled as an inextensible and unshearable elastic rod loaded only at its two ends, with the bulk unaffected by the environment.
Therefore, with $\bp=\bzero$ and $\bl=\bzero$, the theory presented in the previous section can be reduced to the classical elastica.

The configuration of an elastica is completely determined by the signed curvature $\kappa$ of its shape.
An equation governing the curvature of the elastica can be readily derived as follows.
The director components $\{n_1,n_3\}$ of the internal force can be related to the curvature as $n_1=-K\kappa'$ and $n_3 = H - \frac{1}{2}K\kappa^2$.
The former relation is a result of substituting \eqref{eq:constitutive_relation} into \eqref{eq:moment_balance_local}, whereas the latter is a rearrangement of the expression for the Hamiltonian in \eqref{eq:kappa_n_H}.
Using these two relation in $n_1^2+n_3^2=|\bn|^2$ we obtain
\begin{align}\label{eq:shape_equation_dimensional}
    \left(K\kappa'\right)^2 + \left(H-\frac{1}{2} K\kappa^2\right)^2 = |\bn|^2\, .
\end{align}
\begin{figure}[t]
    \centering
    \includegraphics[width=0.8\linewidth]{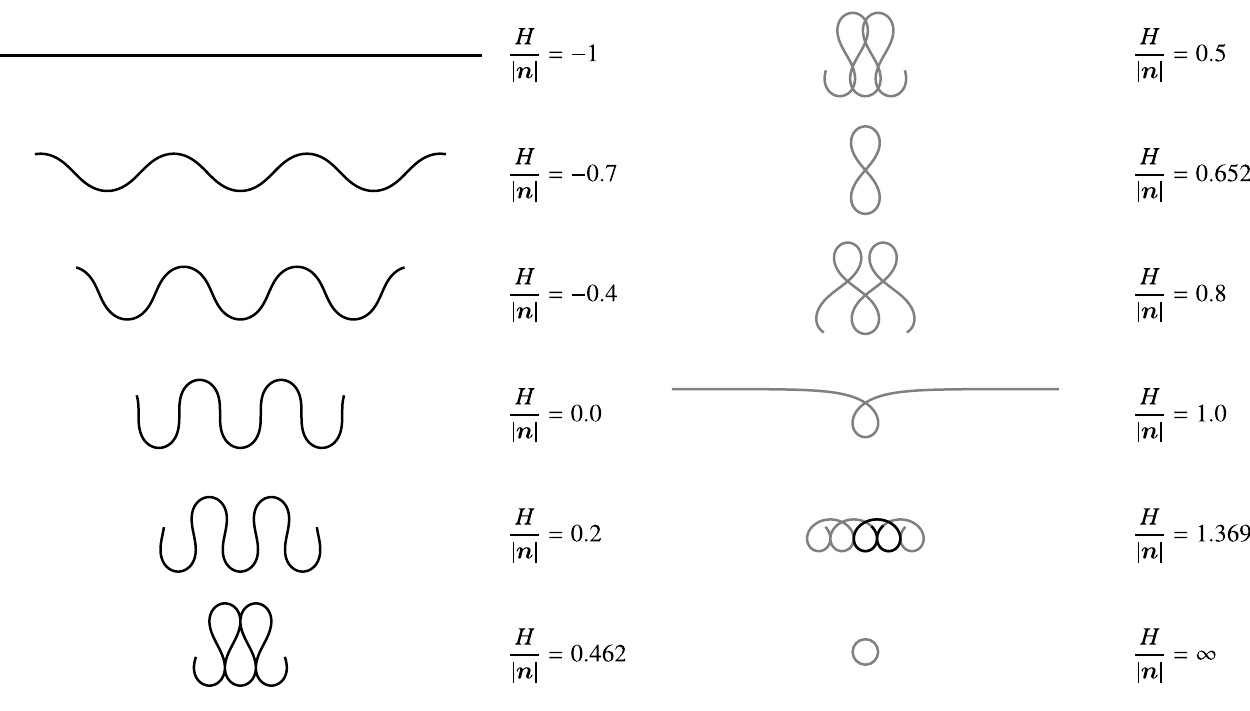}
    \caption{A one-parameter family of solutions for the Euler elastica. Self-avoiding and self-intersecting shapes are shown on the left and the right columns, respectively. The configurations are oriented so that the internal force $\bn$ is horizontal. The segment, highlighted in black, of the mother curve with $H/|\bn| = 1.369$ is the only non-inflectional segment which admits self-contact without self-intersection}
    \label{fig:elastica_non_contact}
\end{figure}
This equation is identical to equation (19) of \cite{singh2019} with $c$ and $B$ there denoted in the present work by $H$ and $K$ respectively.
Since the internal force $\bn$ is conserved in the absence of body forces \eqref{eq:conservation_of_force}, $|\bn|$ is a constant.
Therefore, equation \eqref{eq:shape_equation_dimensional} can be nondimensionalized using the force scale $|\bn|$ and a length scale $\sqrt{K/|\bn|}$ to obtain
\begin{align}\label{eq:shape_equation_nondimensional}
    \left(\kappa'\right)^2 + \left(\frac{H}{|\bn|} - \frac{1}{2}\kappa^2\right)^2 = 1\, .
\end{align}
We conclude from the above that, modulo a boundary condition on $\kappa$, all configurations of an elastica belong to a one-parameter family of curves. 
Equation \eqref{eq:shape_equation_nondimensional} effectively maps the elastica to the motion of a particle in a potential $\tfrac{1}{2}\left(\tfrac{H}{|\bn|}-\tfrac{1}{2}\kappa^2\right)^2$, with kinetic energy $\tfrac{1}{2}(\kappa')^2$, and total energy of $1$ \cite{singh2019}.
The solutions to \eqref{eq:shape_equation_nondimensional} can furthermore be classified as inflectional or non-inflectional,
\begin{subequations}
\begin{alignat}{2}
    \frac{H}{|\bn|} &\in [-1,1)&&\quad \text{inflectional solutions}\, ,\\
    \frac{H}{|\bn|} &\in [1,\infty]&&\quad \text{non-inflectional solutions}\, ,
\end{alignat}
\end{subequations}
while no solutions exist for $\frac{H}{|\bn|}\in (-\infty,-1)$.

Shapes of the elastica for different $H/|\bn|$ values are shown in Fig. \ref{fig:elastica_non_contact}.
Each shape corresponds to a periodic curve that will be referred to as the \emph{mother curve}.

\begin{figure}[h!]
    \centering
    \begin{subfigure}[b]{0.70\textwidth}
        \centering
        \includegraphics[width=\textwidth]{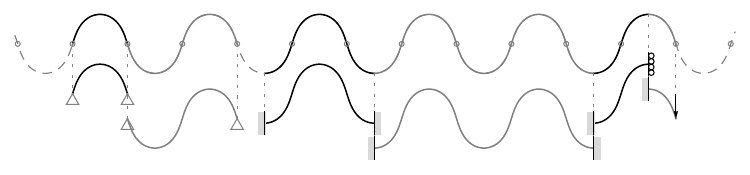}
        % \caption{First subfigure}
        \label{fig:inflectional_mother_curve}
    \end{subfigure}
    
    \begin{subfigure}[b]{0.60\textwidth}
        \centering
        \includegraphics[width=\textwidth]{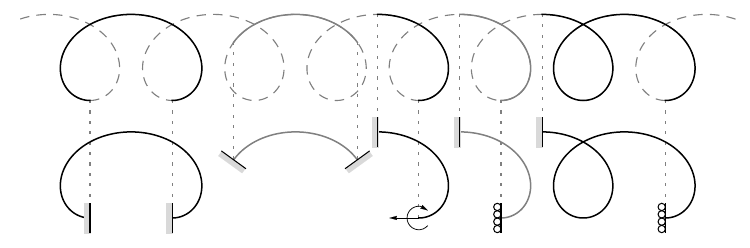}
        % \caption{Second subfigure}
        \label{fig:non_inflectional_mother_curve}
    \end{subfigure}
    \caption{Two mother curves and their segments corresponding to various boundary conditions are shown. The top curve is an inflectional mother curve with $H/|\bn| = -0.3$, while the bottom one is a non-inflectional mother curve with $H/|\bn| = 1.2$.}
    \label{fig:mother_curves}
\end{figure}

As argued in \cite{singh2019}, any solution of the elastica obtained by solving a particular boundary value problem is similar to some segment of the mother curve having the same $H/|\bn|$.
Fig. \ref{fig:mother_curves} shows two mother curves, one with $H/|\bn|=-0.3$ and another with $H/|\bn|=1.2$, along with various elasticae under different boundary conditions that are similar to them.
Consequently, any two configurations with the same value of $H/|\bn|$ share the same mother curve.

We identify three special configurations that we will often allude to in the upcoming analysis.
The first corresponds to the solution with $H/|\bn| = 0$, which we refer to as the \emph{rectangular elastica} \footnote{The name ``rectangular'' is attributed to the fact that the tangent at the point of peak curvature and the inflection point are orthogonal. A rectangular elastica can also be generated by the center of a hyperbola rolling without slipping on a flat surface~\cite{levien2008}.}~\cite{levien2008}.
The second solution corresponds to $H/|\bn|\approx0.462$ which we will call the \emph{critical contact elastica}; this specific ratio is referred to as the \emph{critical value}.
The third one corresponds to $H/|\bn|=1$: a configuration that separates the inflectional solutions from \emph{the} non-inflectional ones, which we will refer to as the \emph{syntractrix}~\cite{levien2008}.  

\section{A scale-invariant contact condition}\label{sec:scale_invariant_contact_condition}
Consider an elastica whose natural or kinematic boundary conditions are continuously (and slowly) varied such that two material points distant in arc-length come close enough in ambient space to establish contact.
Contact between a pair of material points $\{s_1,s_2\}$ is established when their position vectors satisfy the condition
\begin{align}\label{eq:contact_condition}
\br(s_1) = \br(s_2)\, ,\qquad \br'(s_1)\times\br'(s_2) = \bzero\quad\text{such that}\quad s_1,s_2\in(0,L)\quad\text{and}\quad s_1\ne s_2\, .
\end{align}
Detecting this condition for every increment of the varying boundary condition in a systematic way would be cumbersome and inefficient. 
Instead, we establish a condition for the onset of contact based solely on the ratio $H/|\bn|$.

We note from Fig.~\ref{fig:elastica_non_contact} that mother curves corresponding to $H/|\bn|\in(0.462,\infty]$ self-intersect\footnote{We define a point of intersection as a spatial point with $\br(s_1)=\br(s_2)$, such that $s_1\ne s_2$ and $\br'(s_1)\times\br'(s_2)\ne \bzero$}., while those with $H/|\bn|\in[-1,0.462]$ do not.
The mother curve that separates these two classes of configurations is given by the ratio
\begin{align}
    \frac{H}{|\bn|} = 0.46236\, .\label{eq:necessary_condition_for_contact}
\end{align}
As mentioned previously, we refer to this mother curve as the \emph{critical contact curve}.
It also happens to be the only mother curve that contains points of contact (i.e., points that satisfy \eqref{eq:contact_condition}), but does not self-intersect.
Based on this observation, we propose that any inflectional solution of the Euler elastica, regardless of the boundary condition, must necessarily satisfy \eqref{eq:necessary_condition_for_contact} at the onset of contact. 

\begin{figure}[h!]
    \centering
    \includegraphics[width=0.55\linewidth]{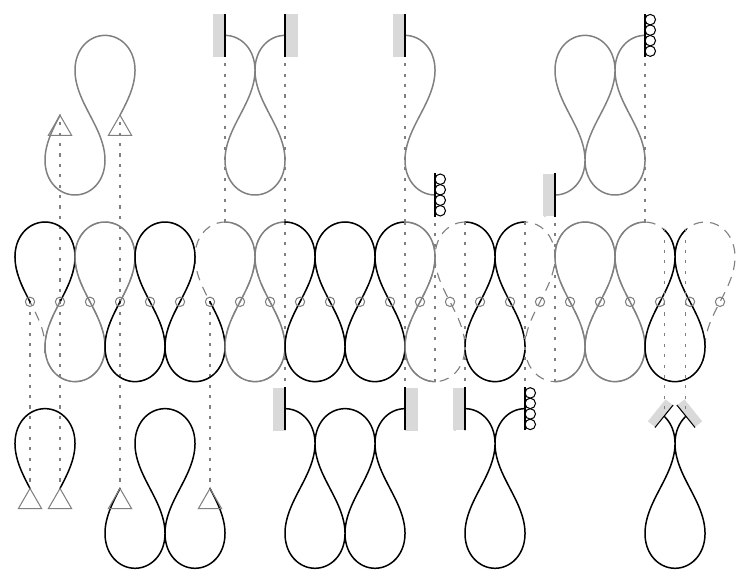}
    \caption{The mother curve of the critical contact elastica, with $H/|\bn|=0.46236$, is shown along with various segments corresponding to different boundary conditions.}
    \label{fig:contact_onset_mother_curve}
\end{figure}

The critical mother curve is shown in Fig.~\ref{fig:contact_onset_mother_curve}, along with various segments of it corresponding to elasticae with different boundary conditions.
We note that condition \eqref{eq:necessary_condition_for_contact} is necessary for the onset of contact in inflectional solutions, but not sufficient.
In other words, the condition \eqref{eq:contact_condition} implies \eqref{eq:necessary_condition_for_contact} for inflectional solutions, but the converse does not hold.
The same is confirmed by the leftmost pinned-pinned configuration below the mother curve in Fig.~\ref{fig:contact_onset_mother_curve}, and the third fixed-roller configuration above it, which share the critical value of $H/|\bn|$, but do not display self-contact.
While \eqref{eq:necessary_condition_for_contact} is not a sufficient condition for contact, it is a much easier criterion compared to \eqref{eq:contact_condition} for detecting the onset of contact in problems where the boundary loads are continuously varying.

A natural follow-up question is whether there exist non-inflectional self-contacting configurations.
While several mother curves that exhibit self-contact in the non-inflectional regime exists, they are always accompanied by points of self-intersection. 
The only exception is a segment of the mother curve corresponding to $H/|\bn| \approx 1.369$ (see Fig.~\ref{fig:elastica_non_contact}) which represents a non-inflectional elastica with self-contact and no self-intersection.

\section{Pinned-Pinned buckled elastica}\label{sec:pinned_pinned_buckled_elastica}
\begin{figure}[t]
    \centering
    \includegraphics[width=0.6\linewidth]{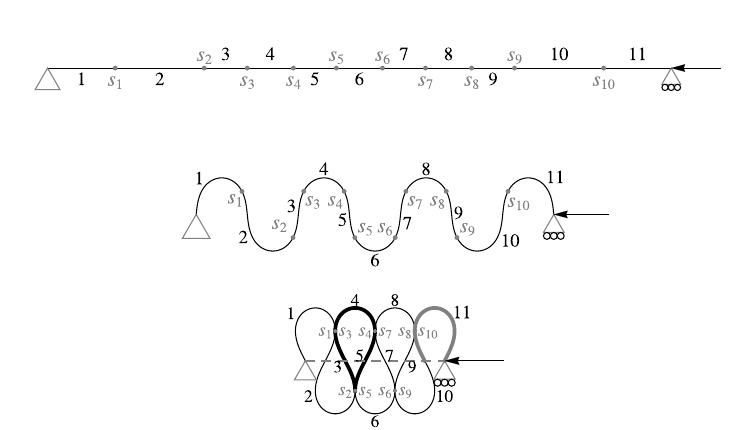}
    \caption{Progression of an axially loaded pinned-pinned elastica buckling into the seventh mode and establishing self-contact. The elastica is divided into several regions, labeled $\{1,2,3\dots\}$, by the contact points $\{s_1,s_2,s_3\dots\}$. The thickened black segment in the self-contacting configuration is referred to as a \emph{lobe}.}
    \label{fig:elastica_pre_contact}
\end{figure}

Here we consider an Euler elastica under a purely axial load $P$ with pinned-pinned boundary conditions (see Fig. \ref{fig:elastica_pre_contact}). 
Let $P_b^{(N)}$ and $P_c^{(N)}$ denote the Euler buckling load and the critical contact load (i.e., the load at the onset of contact), for mode $N$, respectively.
We use condition \eqref{eq:necessary_condition_for_contact}, and the self-similarity of buckling modes, to compute $P_c^{(N)}$, along with the locations of the material points that come into contact at the critical contact load.
Subsequently, we present a formulation to compute post-contact solutions of the buckling modes.
We numerically compute post-contact configurations for modes three to ten.
All numerical computations shown below are performed using numerical continuation of ordinary differential equations as implemented in \texttt{AUTO-07P}~\cite{doedeldldeman2007AUTO}. 

\subsection{Pre-contact}
We begin by stating the boundary value problem governing the pre-contact mechanics of a buckled,  pinned--pinned elastica under a compressive load. 
The equations are nondimensionalized by the length scale $L$ (the length of the elastica) and the force scale $K/L^2$.
\begin{subequations}\label{eq:general_bvp}
    \begin{align}
    r'_1 & = \sin\theta\, ,& r_1(0) &= 0\,,\label{eq:general_bvp_r1}\\
    r'_3 &= \cos\theta\, ,& r_3(0) &= 0\,,\label{eq:general_bvp_r3}\\
    \theta' &= \kappa\, ,& \kappa(0) &= 0\,,\label{eq:general_bvp_theta}\\
    n'_1 &= -\kappa n_3\, ,& -n_1(1)\sin\theta(1) + n_3(1)\cos\theta(1) + P &= 0\,,\label{eq:general_bvp_n1}\\
    n'_3 &= \kappa n_1\, ,& r_1(1) &= 0\,,\label{eq:general_bvp_n3}\\
    \kappa' & = -n_1\, ,& \kappa(1) &= 0\,.\label{eq:general_bvp_kappa}
    \end{align}
\end{subequations}
These equations are obtained by using the planar representations \eqref{eq:planar_representations} of the various fields in \eqref{eq:inextensibility_unshearability} and \eqref{eq:force_moment_balance_local}.
The boundary conditions represent the pinned--pinned supports at the two ends.
Linearization of \eqref{eq:general_bvp} about the straight configuration given by
\begin{align}\label{eq:trivial_config}
    r_1(s) = 0\, ,r_3(s) = s\, ,\theta(s) = 0\, ,n_1(s) = 0\, ,n_3(s) = -P\, ,\kappa(s) = 0\, ,
\end{align}
delivers the following expression for Euler buckling load for mode $N$:
\begin{align}\label{eq:critical_buckling_loads_analytical}
    P_b^{(N)} = N^2\pi^2\, ,\qquad N\in\{1,2,3....\}\, .
\end{align}
At the critical buckling load the trivial solution \eqref{eq:trivial_config} of mode $N$ bifurcates into a non-trivial buckled solution.
No self-contact occurs in the elastica until the load $P$ reaches the critical contact load $P_c^{(N)}$.

\subsection{Onset of contact}
Consider the first buckled mode of an elastica with pinned--pinned boundary conditions.
Let the compressive force $P$ be steadily increased to a point where the ratio $H/|\bn|$ reaches the critical value \eqref{eq:necessary_condition_for_contact}. 
While the first buckled mode does not undergo self-contact at the critical $H/|\bn|$ condition, it remains similar to a segment of the critical contact curve.
The $N$th buckled mode at the critical $H/|\bn|$ is therefore a solution to equations \eqref{eq:general_bvp} with force $P_c^{(N)} = N^2P_c^{(1)}$. This result is a consequence of the self-similarity of the buckling modes~\cite{roman1999}.
Using \eqref{eq:critical_buckling_loads_analytical}, the critical contact load can be related to the buckling load for the $N$th mode as
\begin{align}\label{eq:critical_contact_load_expression}
    P_c^{(N)} = \frac{P_c^{(1)}}{\pi^2}P_b^{(N)}\, .
\end{align}
The numerical value of $P_c^{(1)}=18.047$ for the pinned--pinned boundary condition, and was numerically computed by recording the value of the compressive force $P$ at which the first buckled mode acquires the critical $H/|\bn|$ value.
Relation \eqref{eq:critical_contact_load_expression} is also valid for an elastica with fixed-fixed boundary conditions.

Next, consider the schematic shown in Fig.~\ref{fig:elastica_pre_contact}, with the unloaded configuration at the top, followed by two post-buckling configurations.
The bottom-most configuration depicts the onset of contact.
We now seek to compute the locations of the material points $s_i$ $i\in\{1,2,3,...\}$, that come into contact when $P=P_c^{(N)}$.

The length of each \emph{lobe} (defined by the thickened black segment in Fig.~\ref{fig:elastica_pre_contact}) that forms a repeating unit of any buckled configuration can be computed using scaling arguments. 
The non-dimensional force required to hold a unit length of a full lobe can be computed numerically to be $39.6506$ by solving the governing equations \eqref{eq:general_bvp} with following boundary conditions: $x(0)=0\, ,y(0)=0\, , x(1)=0\, ,y(1)=0,\theta(0)=\pi/2,\theta(1)=3\pi/2$.
Since the length of the lobe scales as the inverse square root of the internal force, the length $l$ of the lobe is given by
\begin{align}\label{eq:length_of_the_loop}
    l = \sqrt{\frac{39.6506}{P_c^{(N)}}}\, .
\end{align}
For the pinned--pinned boundary condition, the $N$th mode contains $N-2$ pairs of material points, i.e., $2(N-2)$ total points, that experience contact at $P=P_c^{(N)}$.
The arc-length coordinate of the $i$th contact point for the $Nth$ buckling mode is given by the following piecewise function
\begin{align}\label{eq:locations_of_contact}
    s_k = \begin{dcases}
               \frac{1}{N}-\alpha\quad &\text{for} \quad k = 1\, , \\
               \frac{1}{N}\left\lceil\frac{k+1}{2}\right\rceil + (-1)^{k+1}\alpha\quad &\text{for} \quad k=2\,\,\text{to}\,\,2N-5\, ,\\
         1 - \left(\frac{1}{N}-\alpha\right)\quad &\text{for} \quad k = 2(N-2)\, ,
    \end{dcases}
\end{align}
where $\alpha = \tfrac{1}{2}(l-\tfrac{1}{N})$, and $\lceil\rceil$ is the ceiling function.
The length $\alpha$ is one-half of the difference between the length of a lobe and the thickened gray region in Fig.~\ref{fig:elastica_pre_contact} (with length $1/N$).

\subsection{Post-contact formulation}
Here we formulate a boundary value problem governing the post-contact response ($P>P_c^{(N)}$) of the buckled elastica incorporating the impenetrability constraint.

For the $N$th mode, we divide the entire domain into $2N-3$ segments separated by $2(N-2)$ unknown contact points.
Let each segment be labeled by the index $J\in\{1,2,\cdots2N-3\}$.
Each segment is then governed by the following set of equations:
\begin{subequations}\label{eq:postcontact_bvp}
    \begin{align}
    (r^J_1)' & = \sin\theta^J\, ,\\
    (r^J_3)' &= \cos\theta^J\, ,\\
    (\theta^J)' &= \kappa^J\, ,\\
    (n^J_1)' &= -\kappa^J n^J_3\, ,\\
    (n^J_3)' &= \kappa^J n^J_1\, ,\\
    (\kappa^J_3)' & = -n^J_1\, .
    \end{align}
\end{subequations}
The conditions connecting the solutions of adjacent segments obtained by enforcing the force and moment jump conditions \eqref{eq:force_moment_jump_conditions} at each contact point. 
At the $k$th contact point, the following jump conditions are enforced
\begin{subequations}\label{eq:continuity_adjacent}
\begin{align}
    \jump{r_1}_k&=0\, ,\label{eq:jump_r1}\\
    \jump{r_3}_k&=0\, ,\label{eq:jump_r3}\\
    \jump{\theta_1}_k&=0\, ,\label{eq:jump_theta}\\
    \jump{n_1}_k&=-P_k\, ,\label{eq:jump_n1}\\
    \jump{n_3}_k&=0\, ,\label{eq:jump_n3}\\
    \jump{\kappa}_k&=0\, ,\label{eq:jump_kappa}
\end{align}
\end{subequations}
where $k\in\{1,2\cdots 2(N-2)\}$.
Conditions \eqref{eq:jump_r1}, \eqref{eq:jump_r3}, and \eqref{eq:jump_theta} enforce the continuity of the position vector and the tangent.
Conditions \eqref{eq:jump_n3} and \eqref{eq:jump_kappa} arise from the frictionless and adhesionless nature of the contact.
Condition \eqref{eq:jump_n1} accounts for the jump in the $\bd_1$ component of the internal force, resulting from an unknown normal contact force $P^k$.
Finally, the pinned--pinned boundary conditions at the two ends of the elastica are given by
\begin{subequations}\label{eq:boundary_conditions}
\begin{align}
   r^1_1(0)&=0\, ,\\
   r_3^1(0)&=0\, ,\\
   \kappa^1(0)&=0\, ,\\
   -n^{2N-3}_1(L)\sin\theta^{2N-3}(1) + n^{2N-3}_3(1)\cos\theta^{2N-3}(1) + P &= 0\, ,\\
   r^{2N-3}_1(1)&=0\, ,\\
   \kappa^{2N-3}(1)&=0\, .
\end{align}
\end{subequations}
Furthermore, we require the equality of position and anti-parallel tangent vectors at all the contact points. 
At the first pair of contact points we have
\begin{subequations}\label{eq:postcontact_bvp_bc_contact_positions_0}
\begin{align}
    r_1(s_1) &= r_1(s_3)\, ,\\
    r_3(s_1) &= r_3(s_3)\, ,\\
    \cos\theta(s_1) &= -\cos\theta(s_3)\, .
\end{align}
\end{subequations}
For the intermediate pairs we have,
\begin{subequations}\label{eq:postcontact_bvp_bc_contact_positions_1}
\begin{align}
    r_1(s_{2n}) &= r_1(s_{2n+3})\, ,\\
    r_3(s_{2n}) &= r_3(s_{2n+3})\, ,\\
    \cos\theta(s_{2n}) &= -\cos\theta(s_{2n+3})
\end{align}
\end{subequations}
where $n=1,2,3,\dots N-4$.
For the last pair of points we have
\begin{subequations}\label{eq:postcontact_bvp_bc_contact_positions_2}
 \begin{align}
    r_1(s_{2(N-3)}) &= r_1(s_{2(N-2)})\, ,\\
    r_3(s_{2(N-3)}) &= r_3(s_{2(N-2)})\, ,\\
    \cos\theta(s_{2(N-3)}) &= -\cos\theta(s_{2(N-2)})\, .
\end{align}   
\end{subequations}
Finally, we specify the the equality of the contact forces between contact points, 
\begin{subequations}\label{eq:postcontact_bvp_bc_force}
\begin{align}
    P_1 &= P_3\, ,\\
    P_{2n} &= P_{2n+3}\, ,\\
    P_{2(N-3)} &= P_{2(N-2)}\, .
\end{align}
\end{subequations}
For mode $N$, the vector of unknown functions is $\{r_1^J,r_3^J,\theta^J,n_1^J,n_3^J,\kappa_3^J\}$, where $J\in\{1,2,\dots2N-3\}$.
This results in $6(2N-3)$ unknown functions.
The unknown parameters are $\{s_i,P_i\}$, where $i=\{1,2,\dots2(N-2)\}$, resulting in $4(N-2)$ unknown parameters.
The total number of unknown functions and unknown parameters is therefore $16N-26$.
Along similar lines, we have $6$ boundary conditions \eqref{eq:boundary_conditions}, $12(N-2)$ jump conditions \eqref{eq:continuity_adjacent}, $3(N-2)$ contact conditions \eqref{eq:postcontact_bvp_bc_contact_positions_0}-\eqref{eq:postcontact_bvp_bc_contact_positions_2}, and $N-2$ conditions on contact forces \eqref{eq:postcontact_bvp_bc_force}, bringing the total to $16N-26$.
The number of boundary conditions is therefore equal to the number of unknown functions plus the number of unknown parameters.

\section{Self-contacting solutions}\label{sec:self_contacting_solutions}
We integrate the boundary value problem derived above using \texttt{AUTO-07P}~\cite{doedeldldeman2007AUTO}, which is a software designed for the numerical continuation of systems of first-order ordinary differential equations.
For a given mode $N$, we first divide the domain of the elastica into segments separated by the points \eqref{eq:locations_of_contact} that will experience contact at the critical contact load $P_c^{(N)}$.
Each segment of this domain is governed by \eqref{eq:postcontact_bvp}.
At the interface of two adjacent sections, we impose the continuity of the solutions of the two elements by prescribing $P_k=0$, and replacing the contact conditions by simply prescribing the numerical values of the arc-length coordinates of the anticipated contact points.

The resulting BVP is then numerically continued from the unloaded straight configuration with  compressive load $P$ as the continuation parameter, and the $N$th bifurcation point is detected.
The trivial solution is then continued along the bifurcating branch (i.e. the post-buckling regime), until the continuation parameter $P=P_c^{(N)}$.
This configuration is then used as a starting solution for the BVP \eqref{eq:general_bvp}-- \eqref{eq:postcontact_bvp_bc_force}, and continued to obtain post-contact solutions with $P$ as the continuation parameter.

As the post-contact configurations evolve, we track the $H/|\bn|$ ratios of each segment of the elastica separated by contact points. 
Here the value of $H$ remains identical for all regions, as discussed in Sec. \ref{sec:nature_of_contact}, but $\bn$ changes from region to region due to the jump in the normal component of the internal force at the contact points.
The $H/|\bn|$ of various regions of the rod is plotted against $P$, which gives us insights into the shapes that each segment of the post-contact configuration acquires.

\subsection{Odd modes}

\begin{figure}[t]
    \centering    
    \includegraphics[width=0.8\linewidth]{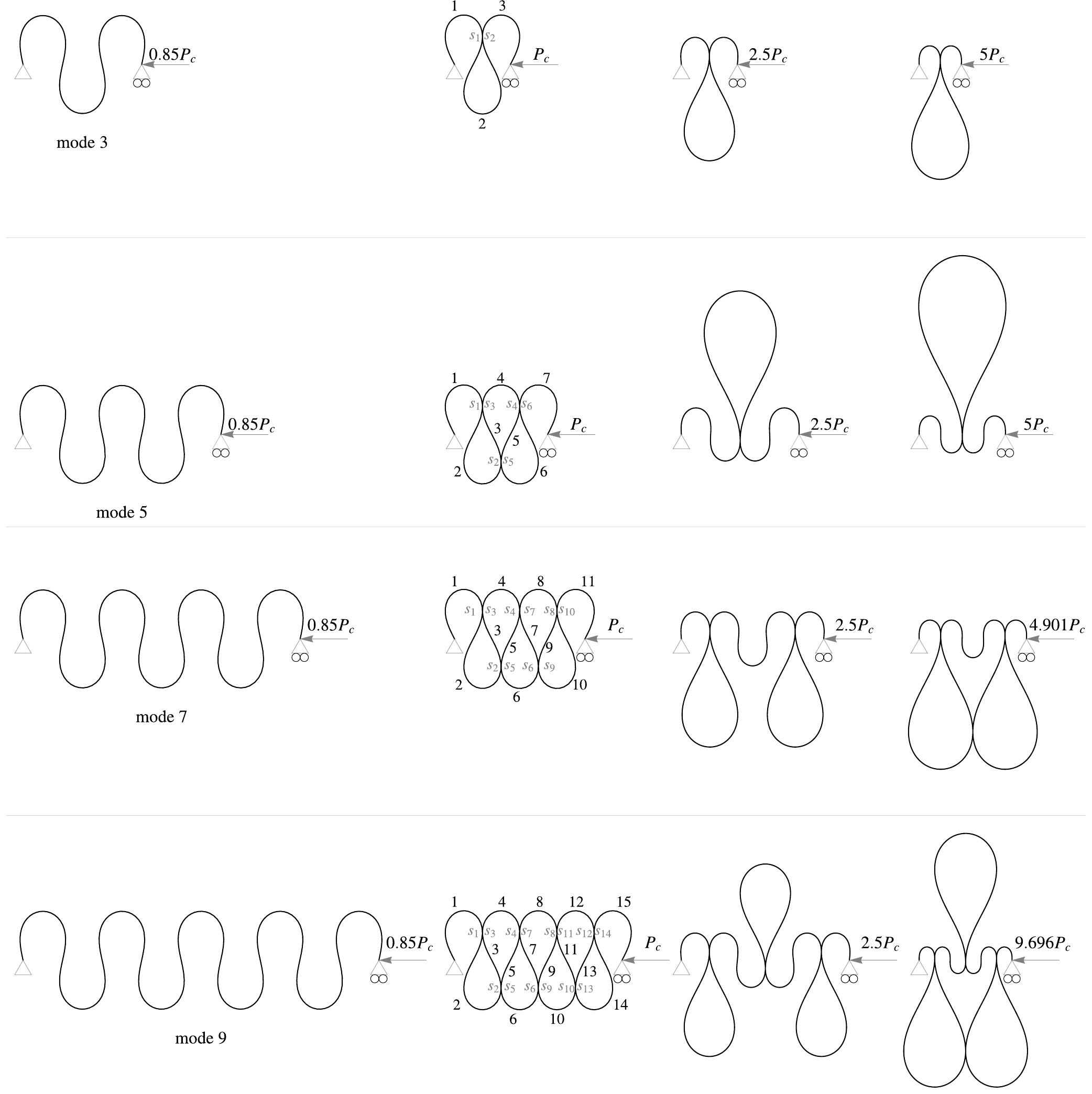}
    \caption{Post-contact configurations for a pinned-pinned elastica buckled into modes 3, 5, 7, and 9. All post-contact shapes consist of two shapes: one is a loop that remains self-similar as the external load is increased, while the rest of the elastica takes on a shape that approaches the shape of a rectangular elastica with increasing compressive force.}
    \label{fig:oddmodes_config}
\end{figure}

Mode one of a buckled elastica does not exhibit contact for any value of the compressive force, and therefore, has not been pursued here.
In mode three, the pair $\{s_1,s_2\}$ of material points comes into contact at a spatial point, as shown in Fig. \ref{fig:oddmodes_config}.
The two points in contact divide the entire elastica into three regions.
As the compressive load $P$ is increased, the region between the two contacting points maintains a self-similar shape with the critical $H/|\bn|$ ratio \eqref{eq:necessary_condition_for_contact}, whereas the remaining two regions asymptotically tend to the shape of a rectangular elastica.
As the compressive load is increased, self-similar loop increases in size, and the material points in contact move toward the terminal ends of the elastica.

In mode five, three pairs, namely $\{(s_1,s_3),(s_2,s_5),(s_4,s_6)\}$, of material points come into contact at three spatial points. 
Continuation of the critical contact solution to post-contact regime leads to a configuration where the reaction force between the pair $(s_2,s_5)$ is adhesive in nature and is therefore inadmissible, whereas the other two pairs exhibit admissible contact.
Computing the post-contact configurations by removing the contact conditions at $(s_2,s_5)$ results in the other two contact pairs, i.e. $(s_1,s_3)$ and $(s_4,s_6)$, turning adhesive.
Admissible configurations are obtained by restoring the contact condition at $(s_2,s_5)$, while removing it from $(s_1,s_3)$ and $(s_4,s_6)$.
This resulting configuration is shown in Fig. \ref{fig:oddmodes_config}.
The shape of this configuration resembles mode three, except for an additional length in the shape that connects the self-similar solution to the two boundaries.
The shape of the curve tends toward a rectangular elastica as the compressive force is increased.
\begin{figure}[h!]
    \centering

    \begin{subfigure}{0.24\textwidth}
        \centering
        \includegraphics[width=\linewidth]{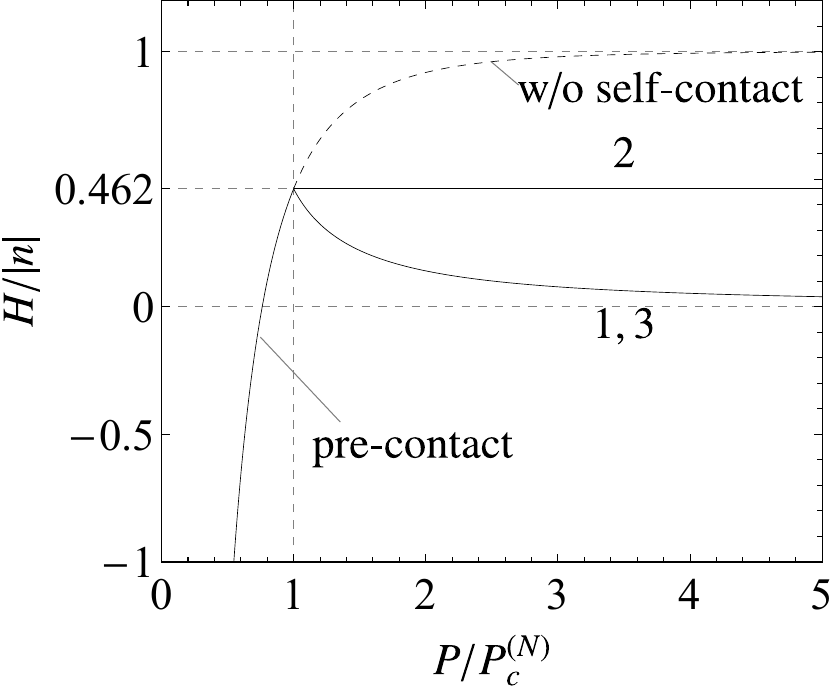}
        \caption{Mode 3}
        \label{fig:mode3HP}
    \end{subfigure}
    \hfill
    \begin{subfigure}{0.24\textwidth}
        \centering
        \includegraphics[width=\linewidth]{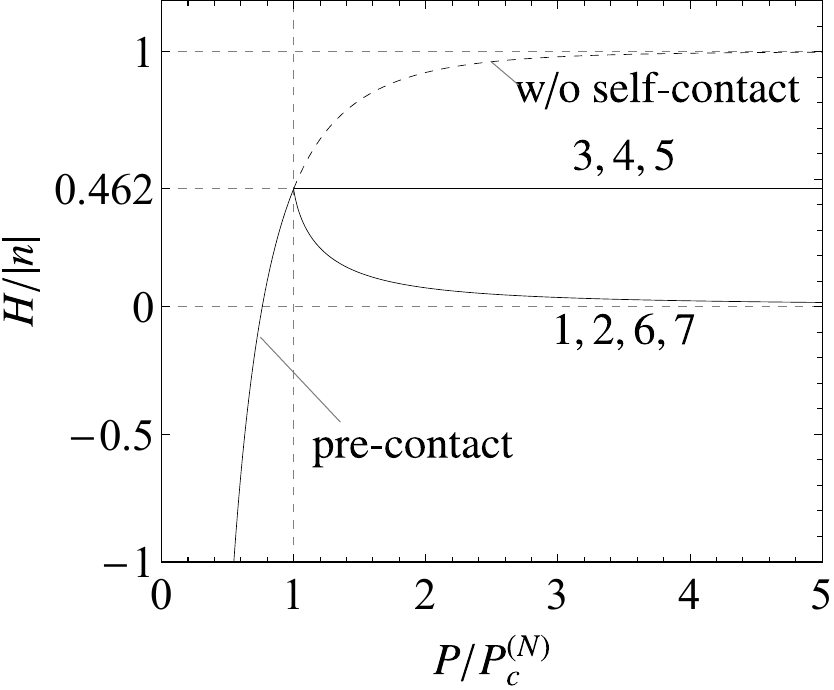}
        \caption{Mode 5}
        \label{fig:mode5HP}
    \end{subfigure}
    \hspace{0cm}
    \begin{subfigure}{0.24\textwidth}
        \centering        \includegraphics[width=\linewidth]{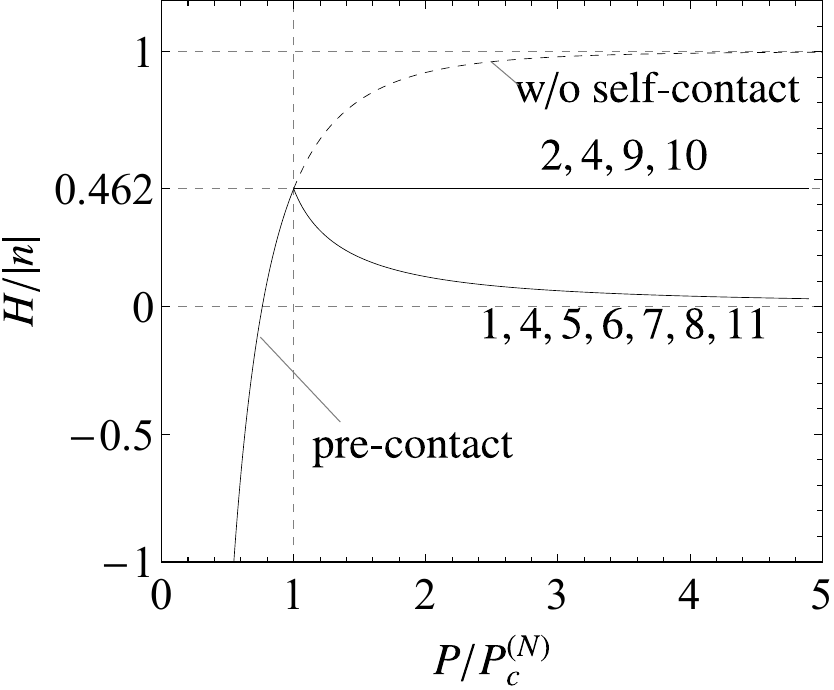}
        \caption{Mode 7}
        \label{fig:mode7HP}
    \end{subfigure}
    \hfill
    \begin{subfigure}{0.24\textwidth}
        \centering
        \includegraphics[width=\linewidth]{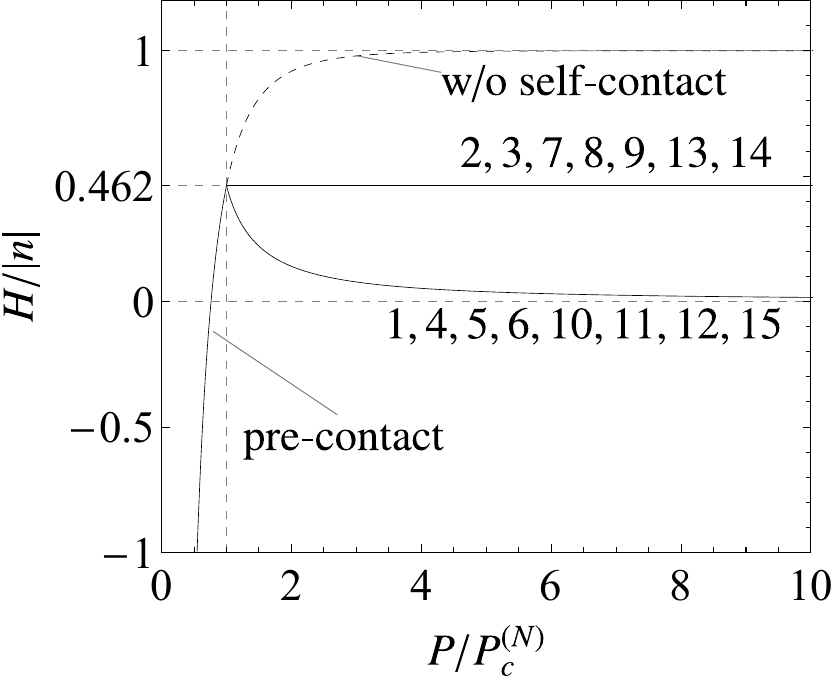}
        \caption{Mode 9}
        \label{fig:mode9HP}
    \end{subfigure}
    \caption{The $H/|\bn|$ vs. $P/P_c$ plots for odd modes. Without accounting for self-contact, all curves tend to $H/|\bn|=1$ (the syntractrix) as the compressive force is increased. With contact incorporated, the curve bifurcates into two branches: one with the self-similar loop, and the other approaching the rectangular elastica where $H/|\bn|\rightarrow 0$.}
    \label{fig:oddHP}
\end{figure}

In mode seven, five pairs, $\{(s_1,s_3),(s_2,s_5),(s_4,s_7),(s_6,s_9),(s_8,s_{10})\}$, of material points come into contact at five spatial points. 
Upon increasing the compressive force, the contact pairs $\{(s_2,s_5),(s_4,s_7),(s_6,s_9)\}$ exhibit adhesive contact, whereas $\{(s_1,s_3),(s_8,s_{10})\}$ undergo admissible contact.
In this case, removing the contact conditions at the adhesive pairs results in an admissible configuration.
The resulting configuration is shown in Fig. \ref{fig:oddmodes_config}.
The configuration appears as two units akin to mode three, connected by a lobe whose shape tends toward a rectangular elastica as the compressive force increases.

In mode nine, seven pairs of material points, $\{(s_1,s_3),(s_2,s_5),(s_4,s_7),(s_6,s_9),(s_8,s_{11}),(s_{10},s_{13}),(s_{12},s_{14})\}$, come into contact at seven spatial points. 
Increasing the compressive load results in a configuration where the pairs $\{(s_4,s_7),(s_8,s_{11})\}$ exhibit adhesive contact, whereas $\{(s_1,s_3),(s_2,s_5),(s_6,s_9),(s_{10},s_{13}),(s_{12},s_{14})\}$ display admissible contact.
To obtain a fully admissible configuration, the contact conditions at pairs $\{(s_2,s_5),(s_4,s_7),(s_8,s_{11}),(s_{10},s_{13})\}$ are removed.
This results in a configuration shown in Fig. \ref{fig:oddmodes_config}, that appears as three units of a mode-three-like configuration connected again by a shape that tends toward a rectangular elastica with increasing load.

In the post-buckling and pre-contact regime of the elastica, the entire shape is characterised by a single value of $H/|\bn|$ as both $H$ and $\bn$ are conserved throughout the length.
Therefore, the shape of the elastica in this regime can be described by one value of $H/|\bn|$ at a given value of $P$.
In the post-contact regime, the internal force $\bn$ is no longer conserved throughout the length since it undergoes jumps at the points of contact.
The same is not true for $H$, which remains conserved along the length of the rod despite contact.

The shape of the elastica in the post-contact regime is, therefore, determined by the individual $H/|\bn|$ ratios of the segments separated by the contact points.
We track and plot in Fig. \ref{fig:oddHP} the ratios $H/|\bn|$ vs the compressive force $P$ normalized by the critical contact load $P_c^{(N)}$ for each mode.
For each mode, the state of the elastica is described by one curve in the pre-contact regime.
The curve then bifurcates into exactly two curves at the onset of contact, i.e. $P=P_c^{(N)}$, for all odd modes studied here.
One branch of the bifurcated curve remains constant at $H/|\bn| = 0.4623$, whereas the other asymptotically tends to $H/|\bn|=0$, i.e. the rectangular elastica.
For the classical elastica, where self-contact is neglected, the curve asymptotically tends to $H/|\bn|=1$, i.e. the syntractrix solution.

\subsection{Even modes}
\begin{figure}[t]
    \centering    
    \includegraphics[width=0.8\linewidth]{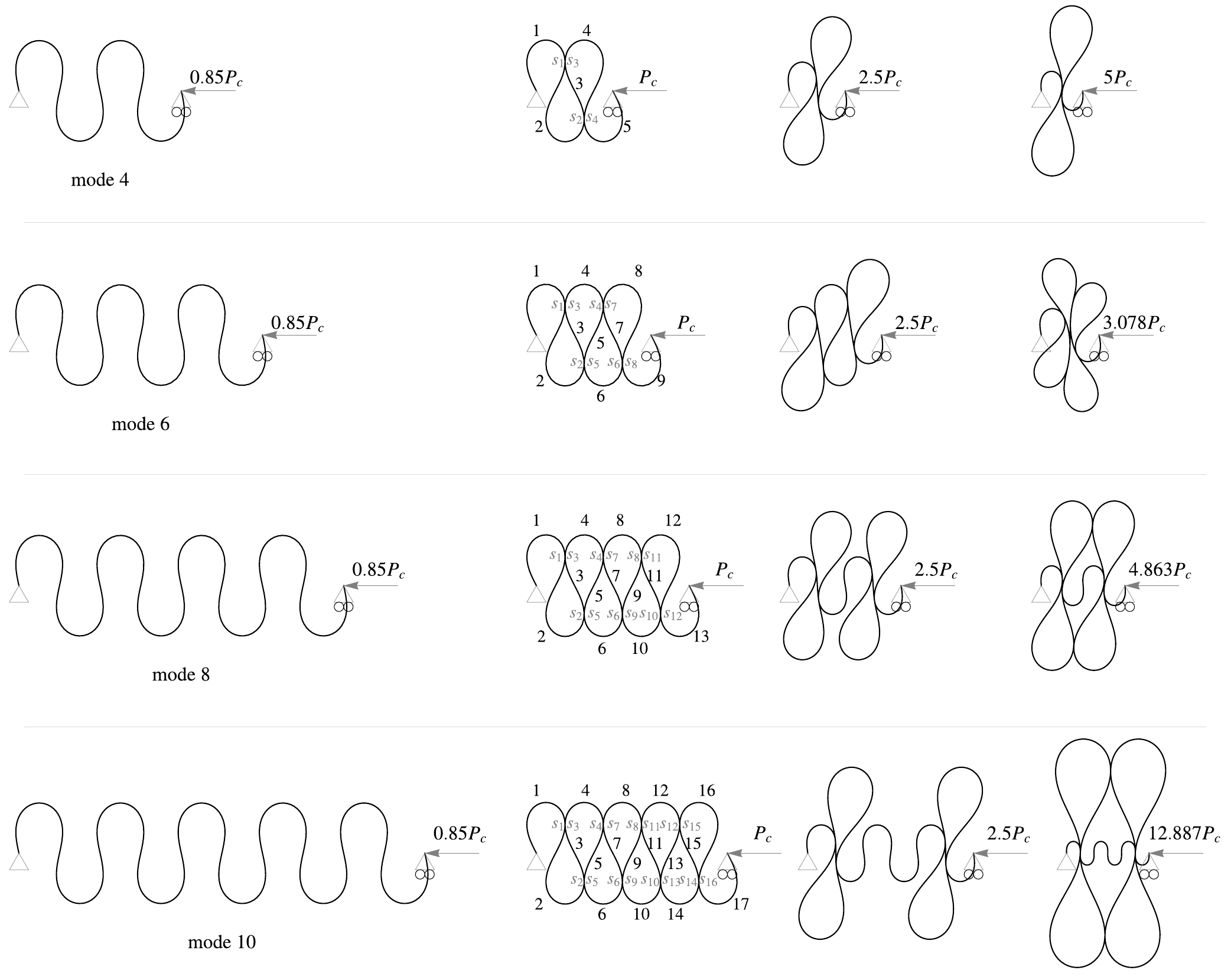}
    \caption{Post-contact configurations for a pinned-pinned elastica buckled into modes 4, 6, 8, and 10. All of the post contact shapes consist of three distinct shapes: a lobe that approaches the self-similar loop ($H/|\bn|=0.4263$) with increasing $P$, a short curve extending from one contact point to another, and the rest of the elastica approaching the rectangular elastica with increasing $P$.}
    \label{fig:evenmodes_config}
\end{figure}

The first even mode, i.e., mode two, does not exhibit contact post-buckling.
In mode four, contact occurs between two pairs of material points, namely $\{(s_1,s_3),(s_2,s_4)\}$ at two spatial points.
As the compressive force is increased post-buckling, the spatial contact points associated with $(s_1,s_3)$ and $(s_2,s_4)$ approach each other, pushing out two lobes in the opposite directions, as shown in Fig. \ref{fig:evenmodes_config}.
\begin{figure}[htbp]
    \centering

    \begin{subfigure}{0.24\textwidth}
        \centering
        \includegraphics[width=\linewidth]{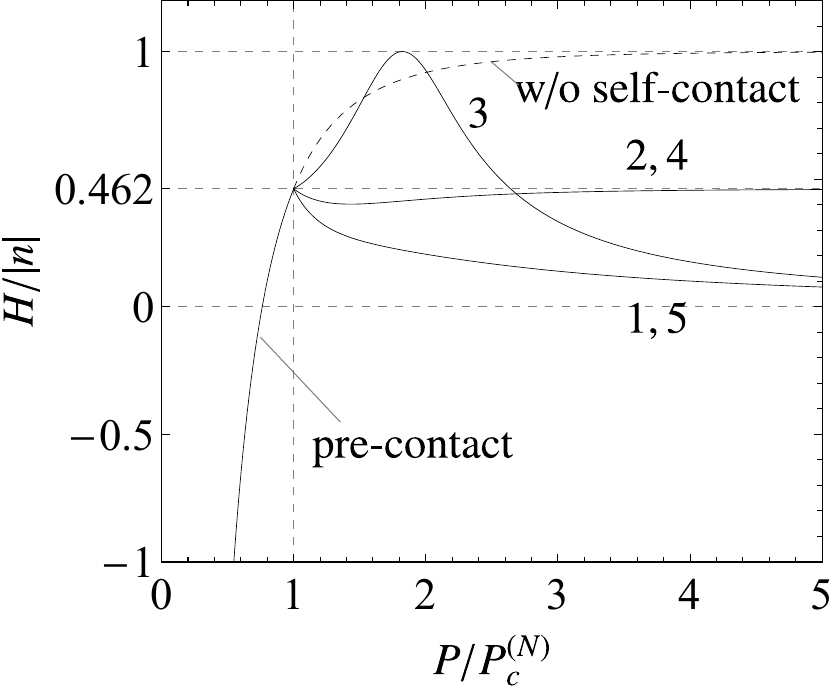}
        \caption{Mode 4}
        \label{fig:mode4HP}
    \end{subfigure}
    \hfill
    \begin{subfigure}{0.24\textwidth}
        \centering
        \includegraphics[width=\linewidth]{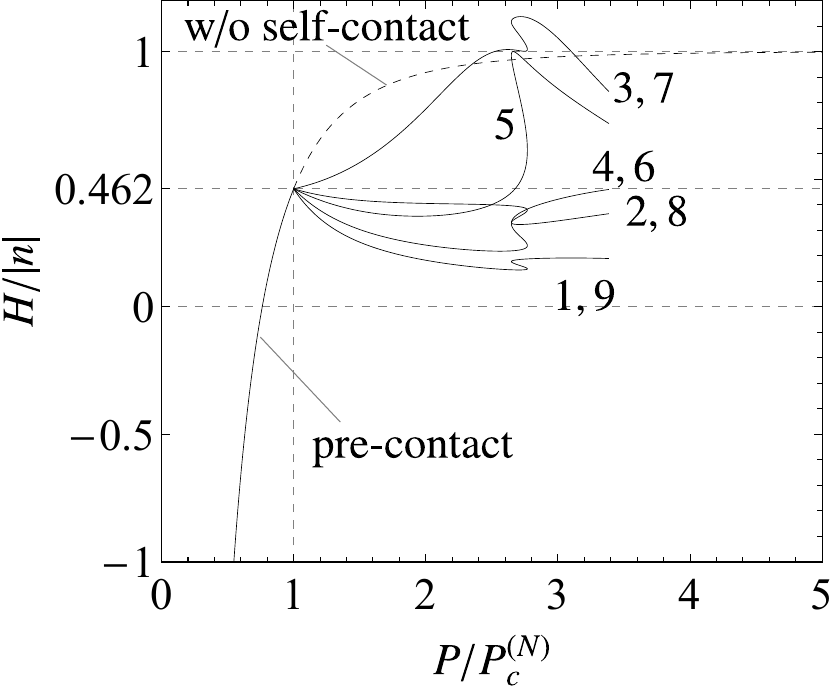}
        \caption{Mode 6}
        \label{fig:mode6HP}
    \end{subfigure}
    \hspace{0cm}
    \begin{subfigure}{0.24\textwidth}
        \centering
        \includegraphics[width=\linewidth]{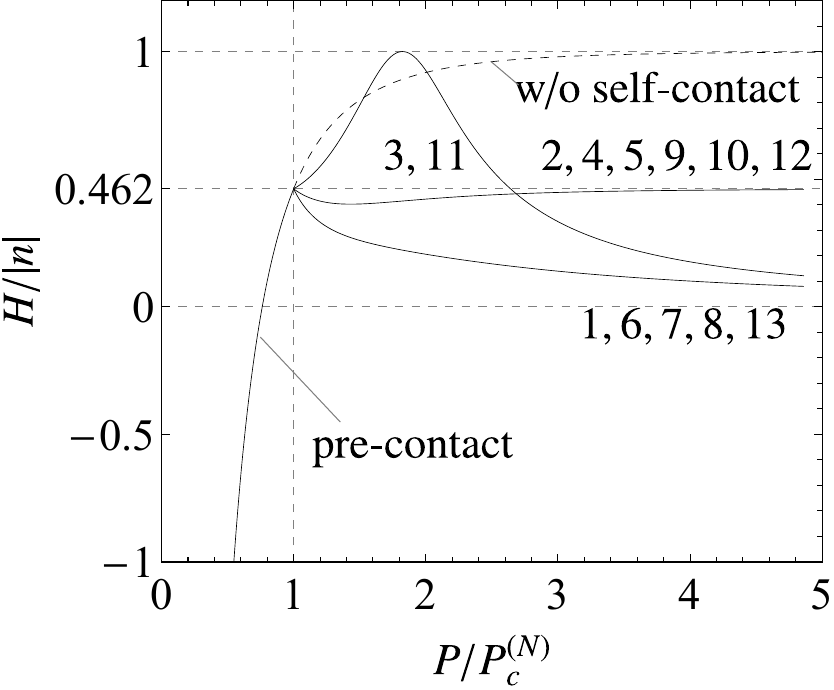}
        \caption{Mode 8}
        \label{fig:mode8HP}
    \end{subfigure}
    \hfill
    \begin{subfigure}{0.24\textwidth}
        \centering
        \includegraphics[width=\linewidth]{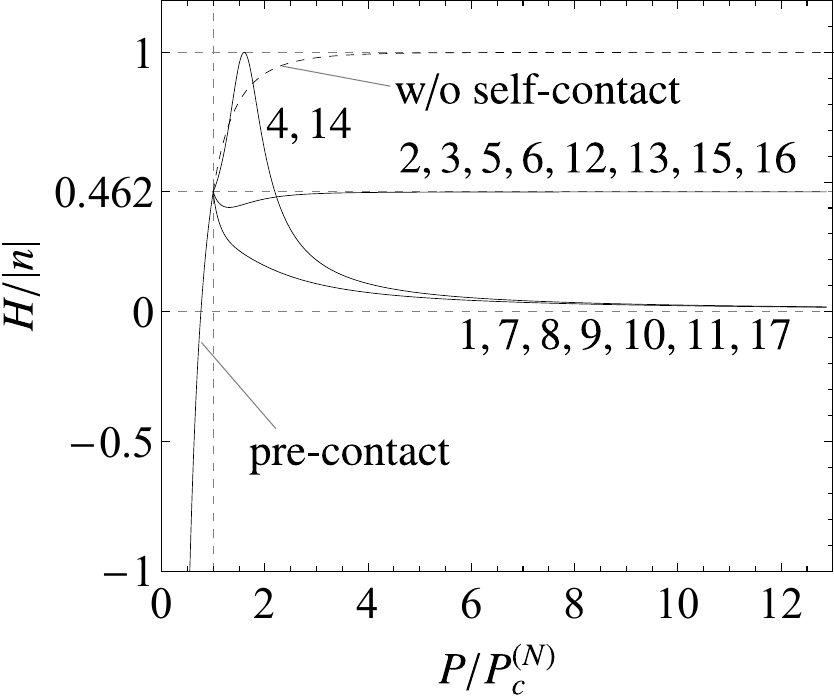}
        \caption{Mode 10}
        \label{fig:mode10HP}
    \end{subfigure}

    \caption{The $H/|\bn|$ vs $P/P_c$ plots of even modes. Without accounting for self-contact, all curves tend to $H/|\bn|=1$ (the syntractrix) as the external load is increased. With contact incorporated, the curve for mode 4, 8, and 10 bifurcates into three branches: one that tends to the self-similar loop with critical $H/|\bn|=0.462$, and the other two approaching the rectangular elastica where $H/|\bn|=0$. Mode 6 deviates from this trend, and does not display the characteristics of the other even modes.}
    \label{fig:evenHP}
\end{figure}
The shape of the outermost regions approaches the rectangular elastica with increasing compressive force.
The shape of the two lobes tends to the critical loop, with the $H/|\bn|$ ratio given by \eqref{eq:necessary_condition_for_contact}.

Mode six appears to be the most peculiar among the even modes.
It displays four pairs of contact: $\{(s_1,s_3),(s_2,s_5),\\(s_4,s_7),(s_6,s_8)\}$.
As the compressive force is increased, four lobes are formed, as shown in the Fig. \ref{fig:evenmodes_config}, and the entire configuration is pushed against itself.
Beyond a load of $P \approx 3.078 P_c$, the middle region exhibits self-penetration in a complex fashion. 
Among all the modes that we've investigated in this work, mode six is the only case where such penetration occurs.

Mode eight displays contact between six pairs of material points $\{(s_1,s_3),(s_2,s_5),(s_4,s_7),(s_6,s_9),(s_8,s_{11}),(s_{10},s_{12})\}$.
As the compressive force is increased, the contact pairs $\{(s_2,s_5),(s_8,s_{11})\}$ become adhesive, whereas $\{(s_1,s_3),(s_4,s_7),\\(s_6,s_9),(s_{10},s_{12})\}$ remain be admissible.
We remove the contact at pairs $\{(s_2,s_5),(s_4,s_7),(s_{10},s_{12})\}$, and obtain a fully admissible configuration, as shown in Fig. \ref{fig:evenmodes_config}.
The resulting configuration appears as two mode-four-like units connected to one another by a shape that tends toward a rectangular elastica as the compressive force is increased.

Finally, mode ten displays eight pairs of material points, namely $\{(s_1,s_3),(s_2,s_5),(s_4,s_7),(s_6,s_9),(s_8,s_{11}),(s_{10},s_{13}),\\(s_{12},s_{15}),(s_{14},s_{16})\}$, establishing contact.
The pairs $\{(s_2,s_5),(s_6,s_9),(s_8,s_{11}),(s_{12},s_{15})\}$ result in adhesive contact, whereas the remaining pairs $\{(s_1,s_3),(s_4,s_7),(s_{10},s_{13}),(s_{14},s_{16})\}$ are admissible.
The pairs $\{(s_4,s_7),(s_6,s_9),\\(s_8,s_{11}),(s_{10},s_{13})\}$ are then released to obtain a fully admissible configuration.
The resulting configuration qualitatively appears as two mode-four-like units connected by a shape that approaches the rectangular elastica as the compressive force increases.

As done for the odd modes, we plot the $H/|\bn|$ vs. $P/P_c^{(N)}$ curves for the even modes, as shown in Fig.\ref{fig:evenHP}.
The single curve from the pre-contact regime bifurcates into three curves for mode four, eight, and ten.
One of these curves tends toward the critical contact state defined by \eqref{eq:necessary_condition_for_contact}.
The second curve initially rises toward $H/|\bn| = 1$ (the syntractrix solution) and then approaches $H/|\bn|=0$ (the rectangular elastica). 
Finally, the third curve approaches the rectangular elastica immediately after the onset of contact.

Mode six appears more peculiar than the other even modes. 
Its pre-contact $H/|\bn|$ curve bifurcates into four curves with no clear tendencies to approach limiting shapes.
The continuation process is terminated at the point where secondary contact occurs.
We suspect that while this seemingly anomalous behavior may manifest in higher modes, especially modes that are integer multiples of six.
A study of mode 12 and 18 may confirm or deny this hypothesis.

\subsection{Mode 8 and 9: additional configurations}
In the previous two subsections, whenever an adhesive contact pair was detected, a certain set of contacts was released in order to obtain a fully admissible configuration. 
Our choices of which contact pairs to release were based on trial and error, instead of a definite criterion.
In this process, we found more than one admissible configuration for mode eight and nine by releasing different sets of contact points.
Here we present configurations different from the ones presented in Fig.~\ref{fig:oddmodes_config} and \eqref{fig:evenmodes_config} for mode eight and nine, demonstrating the existence of multiple configurations for the same compressive force.
\begin{figure}[h!]
    \centering
    \includegraphics[width=0.75\linewidth]{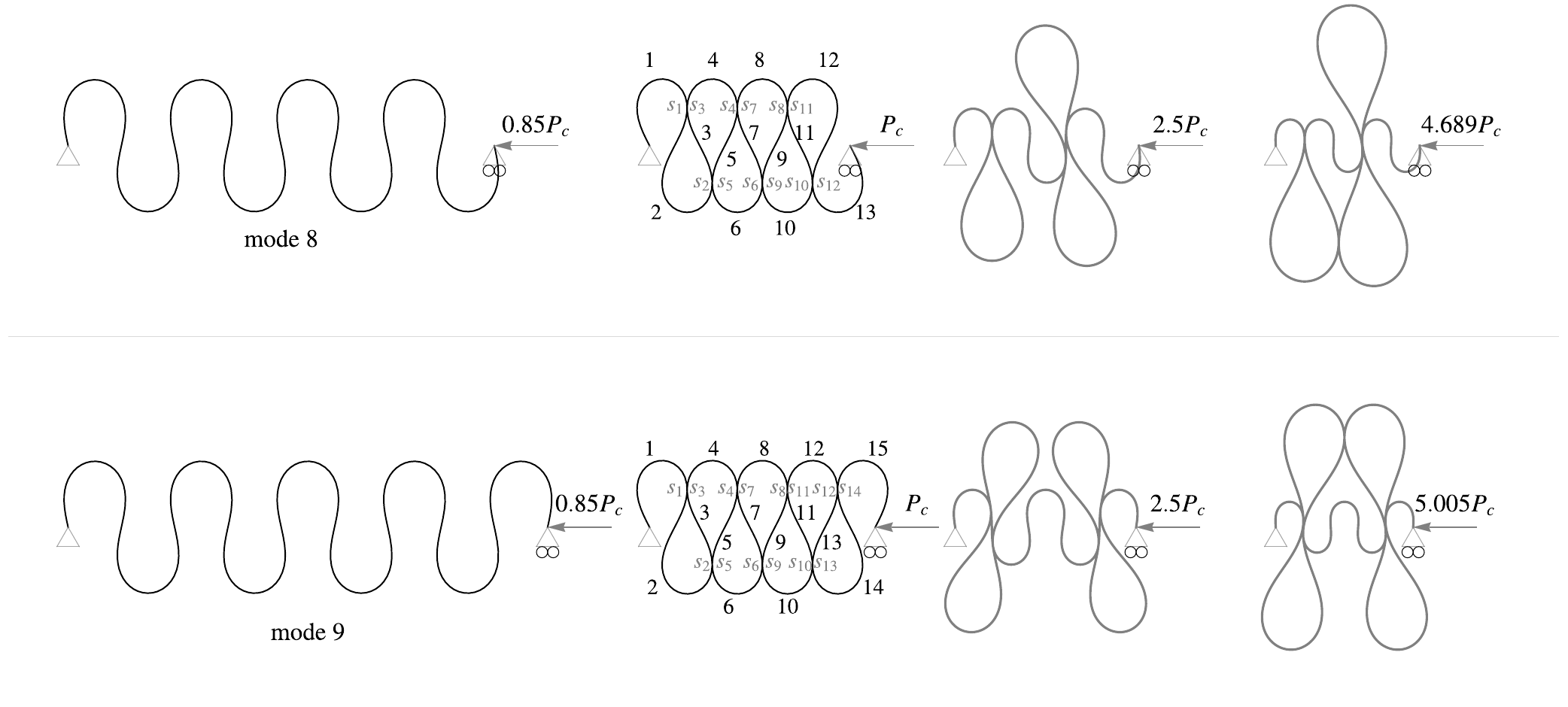}
    \caption{Additional configurations for modes 8 and 9, obtained by removing contact between specific pairs of contact points. In mode 8, the pairs $\{(s_4,s_7),(s_6,s_9)\}$ were removed, whereas $\{(s_4,s_7),(s_6,s_9),(s_8,s_{11})\}$ were removed for mode 9.}
    \label{fig:additionalconfig}
\end{figure}

\begin{figure}[htbp]
    \centering
    \begin{subfigure}{0.35\textwidth}
        \centering
        \includegraphics[width=
        \linewidth]{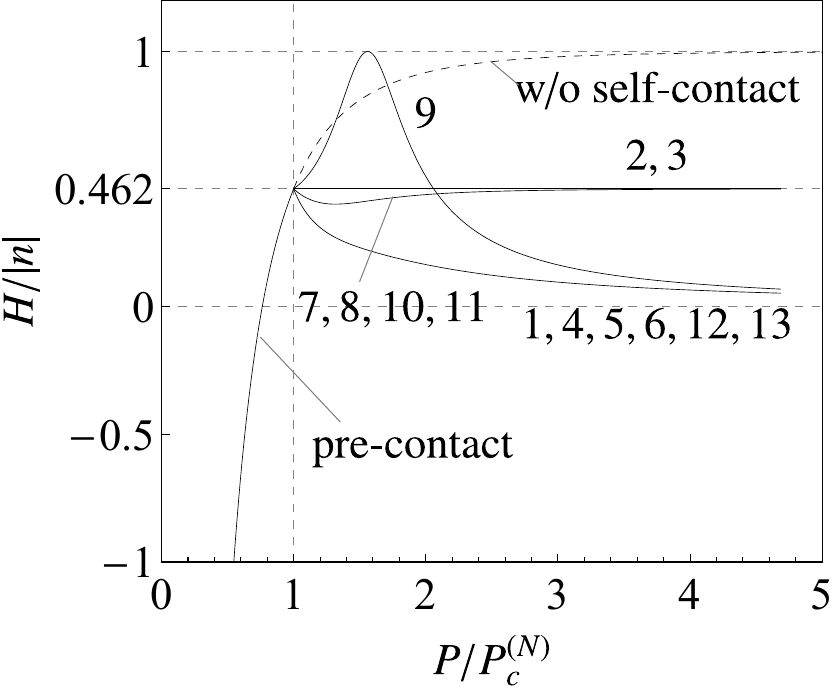}
        \caption{Mode 8}
        \label{fig:mode8HPadditional}
    \end{subfigure}
    \hspace{0cm}
    \begin{subfigure}{0.35\textwidth}
        \centering
        \includegraphics[width=\linewidth]{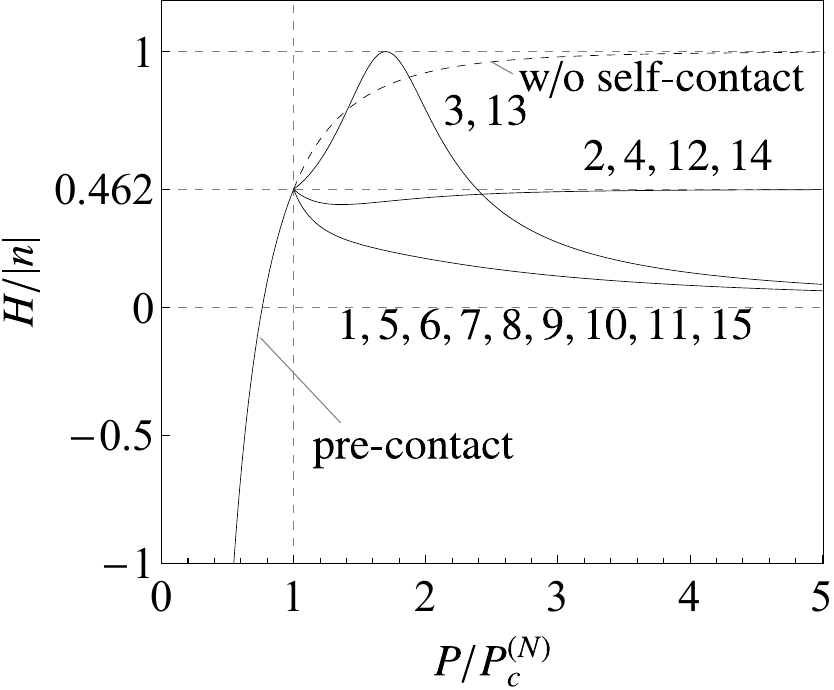}
        \caption{Mode 9}
        \label{fig:mode9HPadditional}
    \end{subfigure}
    \caption{The $H/|\bn|$ vs. $P/P_c$ plots of the additional configurations obtained for mode 8 and 9. Both plots exhibit characteristics of the odd and even modes.}
    \label{fig:HPadditional}
\end{figure}

For mode eight, a fully admissible configuration is also obtained by removing the contact at pairs $\{(s_4,s_7),\\(s_6,s_9)\}$.
The resulting shape is noteworthy because, unlike the configuration for mode eight shown in Fig. \ref{fig:evenmodes_config}, it is a combination of a mode-three-like unit connected to a mode-four-like solution by a shape that approaches the rectangular elastica as the compressive force increases.
This is supported by the $H/|\bn|$ vs. $P/P_c^{(N)}$ plot shown in Fig.\ref{fig:mode8HPadditional}, which exhibits characteristics of both even and odd modes.
% At $P\approx 4.689P_c$, secondary contact is established between the lobes.

For mode nine, a second admissible configuration is obtained by removing the contact at $\{(s_4,s_7),(s_6,s_9),(s_8,s_{11})\}$.
The resulting shape is shown in Fig. \ref{fig:additionalconfig}.
Despite being an odd mode, the resulting post-contact shape differs qualitatively from its counterpart in Fig.~\ref{fig:oddmodes_config} and appears to be composed of two mode-four-like units connected by a shape approaching the rectangular elastica.
The lobes of the two units here approach the self-similar solution with the critical $H/|\bn|$ ratio given by \eqref{eq:necessary_condition_for_contact}. 
The $H/|\bn|$ plot for the resulting shape is shown in Fig. \ref{fig:mode9HPadditional}.

The existence of these additional configurations for modes eight and nine leads us to believe that higher modes (i.e. mode 10 and above) may also admit multiple configurations at the same contact load.
The release of appropriate contact pairs may also lead to configurations that qualitatively resemble mode six, as well as similar $H/|\bn|$ vs $P/P_c^{(N)}$ curves.
However, in the absence of a clear criterion to identify the contact pairs whose release would lead to admissible configurations, our attempts to find additional configurations for mode 10 did not succeed.

% Secondary contact between the lobes happens around $P\approx 5P_c^{(9)}$.

\section{Line contact in odd and even modes}\label{sec:line_contact}
In the computations presented so far, we have only considered point-contact between various points on the elastica.
It is reasonable to imagine that if any of these configurations are pushed hard enough, they may eventually lead to line-contact, wherein two segments of the elastica, instead of just two points, are in continuous contact.
Here, we argue that line-contact cannot exist for the post-contact configurations of the odd modes shown in Fig.~\ref{fig:oddmodes_config} for any finite value of the compressive force.
\begin{figure}[h!]
    \centering
    \begin{subfigure}{0.35\textwidth}
        \centering
        \includegraphics[width=
        \linewidth]{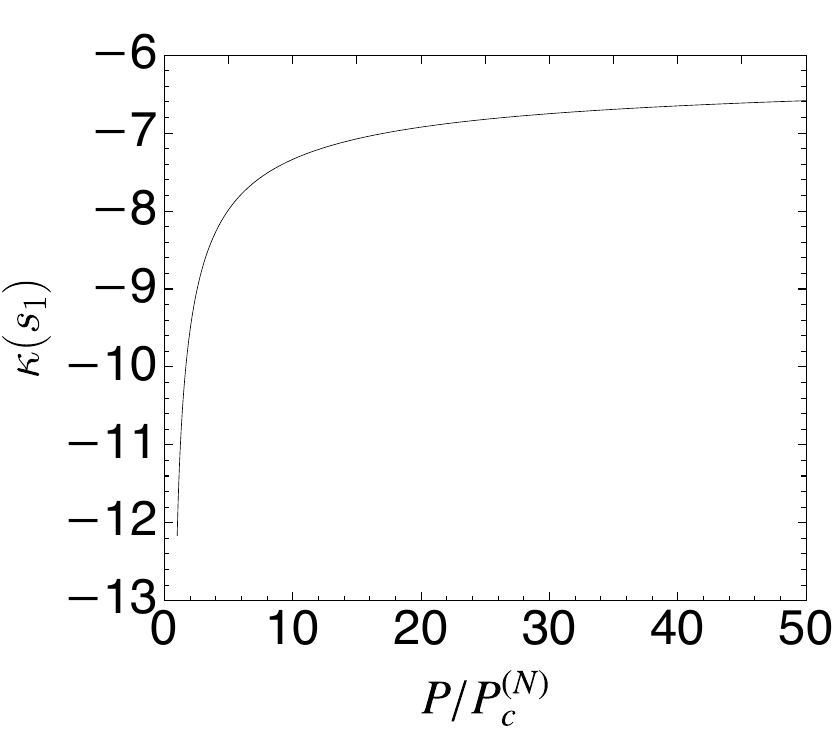}
        \caption{Mode 3}
        \label{fig:mode3curvature}
    \end{subfigure}
    \hspace{0cm}
    \begin{subfigure}{0.35\textwidth}
        \centering
        \includegraphics[width=\linewidth]{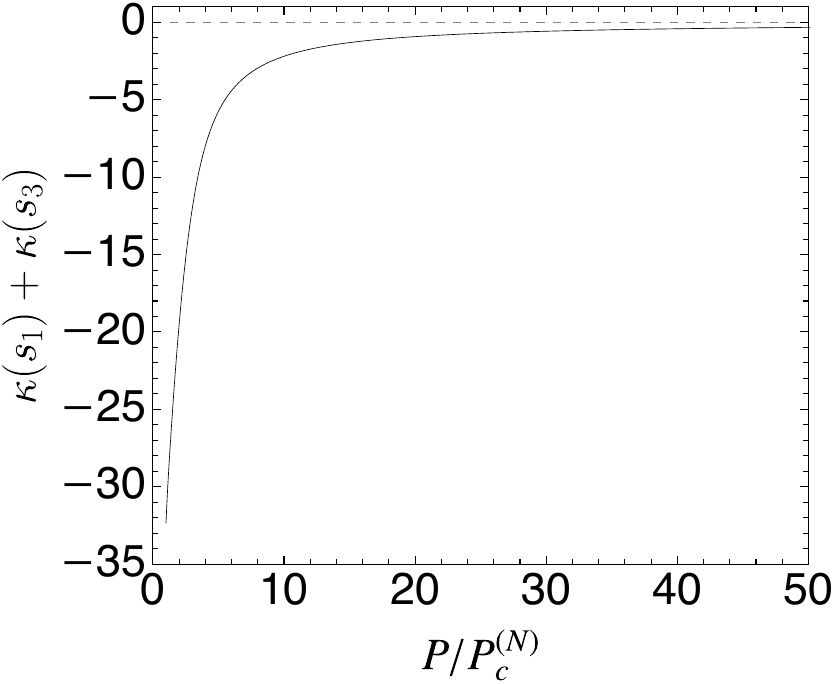}
        \caption{Mode 4}
        \label{fig:mode4curvature}
    \end{subfigure}
    \caption{(a) The signed curvature of the contact point $s_1$ of mode 3 as a function of $P/P_c$. (b) Sum of the signed curvatures of points $s_1$ and $s_2$ of mode 4 as a function of $P/P_c$.}
    \label{fig:curvatureplot}
\end{figure}

Let us consider the configurations shown in Fig. \ref{fig:oddmodes_config}.
Since each of these configurations comprises repeated units of shapes of mode three, we will present our argument for mode three alone.
If a point of contact in mode three were to transition into line contact, then, by symmetry, the line contact region would have to be straight (i.e. $\kappa = 0$) and purely vertical, implying $\bn\cdot\bd_3=  n_3 = 0$ along that segment. 
This further implies that the Hamiltonian in the line segment must be zero making $H/|\bn|=0$ for the entire configuration.
Since the Hamiltonian is conserved throughout the length of the elastica, this is incompatible with the self-similar lobe in the configuration, whose $H/|\bn|$ ratio is fixed by \eqref{eq:necessary_condition_for_contact}. 
The only way to resolve this contradiction is to conclude that a region of line contact cannot exist in configurations shown in Fig.~\ref{fig:oddmodes_config} which comprise repeated units of mode three.
This can be numerically verified in the curvature vs. $P/P_c^{(3)}$ plot for mode three in Fig.~\ref{fig:mode3curvature}.
\begin{figure}
    \centering
    \includegraphics[width=0.85\linewidth]{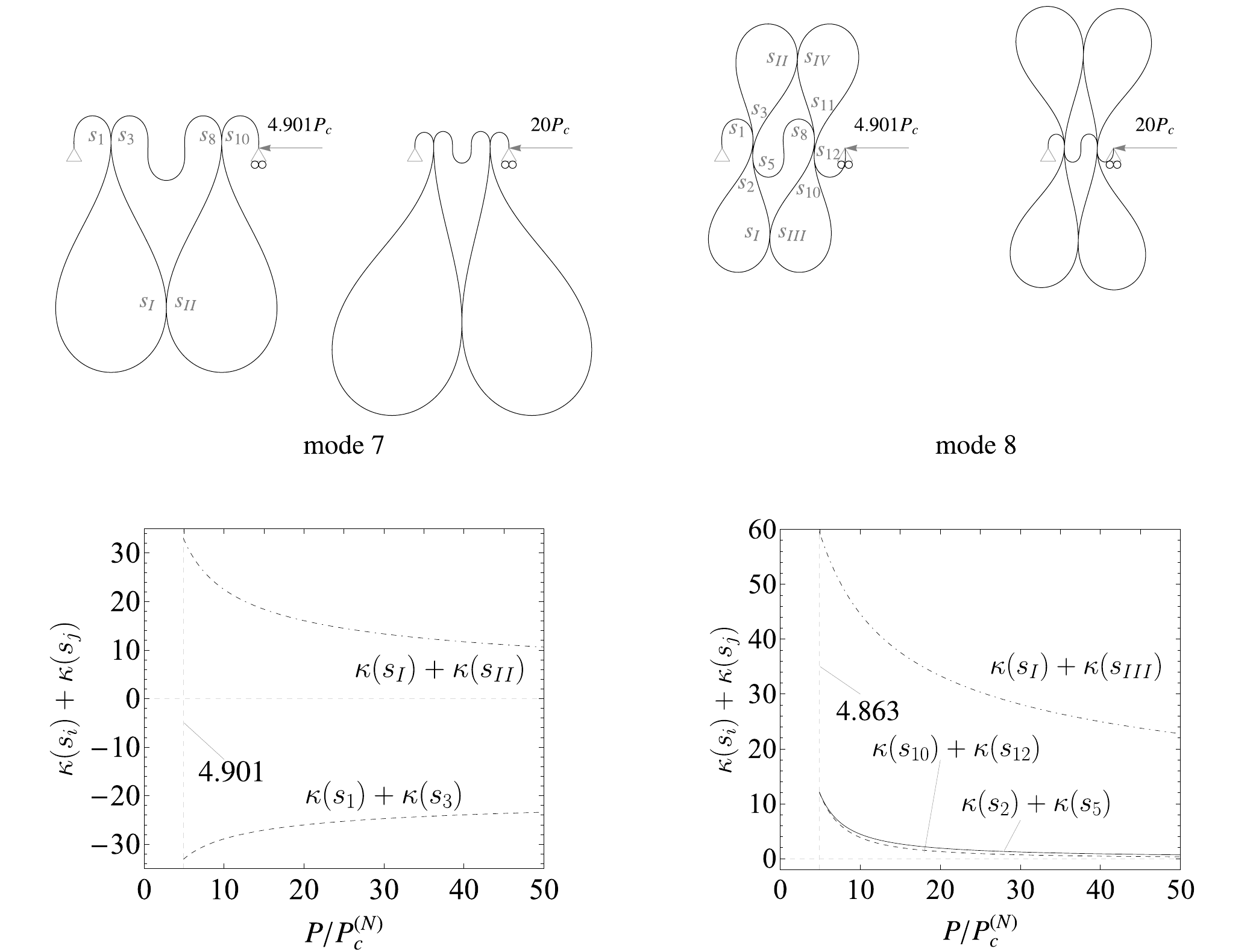}
    \caption{Configurations beyond the secondary contact for mode 7 (top left) and  mode 8 (top right). The sum of the signed curvatures of various contact pairs for mode 7 (bottom left) and mode 8 (bottom right).}
    \label{fig:secondarycontact}
\end{figure}

Our argument for the odd modes does not work for the even modes, as the latter are not symmetric.
However, we numerically investigate the possibility of a point contact transitioning to a line contact in mode four, since modes eight and ten contain mode-four-like shapes.
It was observed that with increasing compressive force, the two pairs of contact points, i.e. $\{(s_1,s_3),(s_2,s_4)\}$ approach each other in ambient space, while the two lobes approach the critical $H/|\bn|$ ratio given by \eqref{eq:necessary_condition_for_contact}.
Numerical simulations show that the point contacts do not transition into line contact until $P=50 P_c^{(4)}$, where the criterion to check for line contact was the vanishing of the sum of the signed curvatures at the contact pairs.
The same is plotted in Fig. \ref{fig:mode4curvature}, where the sum of the signed curvatures approach zero only asymptotically.

As the various post-contact configurations for higher modes evolve with increasing compressive force new contact points may form between different parts of the rod.
We will refer to these new points as points of secondary contact.
This occurs in mode seven (at $P\approx 4.9 P_c^{(7)}$), mode nine (at $P\approx 9.7 P_c^{(7)}$), mode eight (at $P\approx 4.9 P_c^{(7)}$), and mode ten (at $P\approx 12.9 P_c^{(7)}$).
In order to explore the possibility of line contact, we compute configurations for one even and one odd mode (mode seven and mode eight), beyond the onset of secondary contact.
Fig.~\ref{fig:secondarycontact} show the resulting configurations, and the sum of the signed curvatures at all contact points. 
For line contact to develop, the sum of the signed curvatures must vanish.
We ran both configurations up to 50 times the critical contact load, however, no point contact was seen to transition into line contact.

\section{Conclusion}\label{sec:conclusion}
We have considered a buckled elastica compressed to an extent that different parts of it come into contact.
A scale-invariant criterion \eqref{eq:necessary_condition_for_contact} involving two integrals associated with the classical elastica was proposed to establish the onset of contact.
As a case study, the buckling of an elastica with pinned--pinned boundary conditions was considered.
The critical contact load $P_c^{(N)}$ for a given $Nth$ mode, i.e., the force at which contact in a buckled elastica commences, was related to the corresponding buckling load \eqref{eq:critical_contact_load_expression}.
The arc-length coordinates of the material points for the $N$th mode that come into contact at $P=P_c^{(N)}$ were computed \eqref{eq:locations_of_contact}.
Finally, a full boundary value problem governing the post-contact mechanics of a buckled elastica was derived.

The boundary value problem was solved numerically using numerical continuation.
Post contact configurations for odd and even modes, from mode three through mode ten, were studied separately due to their qualitatively different morphology.
It was shown that, in general, post-contact configurations of odd modes comprise exactly two characteristic shapes, as identified by tracking the $H/|\bn|$ ratio as a function of the compressive force.
One shape corresponds to the critical $H/|\bn|$ value, while the other approaches the rectangular elastica ($H/|\bn|\rightarrow 0$) as $P$ increases.
The modes five, seven, and nine appeared as modular repetitions of mode-three-like shapes.

For even modes, with the exception of mode six, the post-buckling shapes consisted of three shapes. 
Two of these shapes approached the rectangular elastica with increasing compressive force, while one approached the critical self-similar loop corresponding to the critical $H/|\bn|$ value \eqref{eq:necessary_condition_for_contact}.
Mode six stood out as an exception, where the post-contact shape comprised four different shapes represented by four curves in the $H/|\bn|$ vs. $P/P_c^{(6)}$ plots, with no clear evolution patterns for any of them apparent.
Furthermore, mode eight and ten appeared as modular repetitions of mode-four-like shapes.

Our initial computations revealed that the post-contact configurations for several modes admit adhesive contact at various points. 
To obtain fully admissible configurations, we removed several contact constraints by trial and error to obtain configurations that do not contain any adhesive joints.
In this process, we discovered two admissible configurations each for modes eight and nine.
The resulting configuration for mode eight appeared to incorporate a mode-three-like and a mode-four-like shape, while mode nine appears to be comprised of purely mode-four-like shapes.
The existence of these additional configurations indicates that perhaps other admissible configurations may arise for higher modes upon the removal of different sets of contact pairs. A systematic way of determining the pairs of contact points whose removal leads to admissible configurations remains an open question.

Finally, we considered the possibility of point contact transitioning into line contact.
We presented a theoretical argument, based on the conservation of the Hamiltonian, showing that line contact is impossible in the symmetric odd modes configurations shown in Fig.~\ref{fig:oddmodes_config}.
For even modes, we provided numerical evidence that line contact does not develop as the compressive force increases.
Occurrence of secondary contacts in mode seven and eight was observed at certain values of the compressive loads.
Post secondary contact configurations for both these modes were computed.
No line contact was observed in the post-secondary contact configurations upto the loads investigated.

\section{Acknowledgments} 
K.S. acknowledges that part of this work was conducted during his postdoctoral tenure at IIT Gandhinagar. 
H.S. and P.P. are grateful to the Science and Engineering Research Board (SERB) for financial support under grant SRG/2023/000079 awarded to H.S.
% \bibliography{refs_master_url}

%

\end{document}